\documentclass[10pt, a4paper]{article}

\usepackage{subfigure}
\usepackage{graphicx}

\usepackage{amsfonts, amssymb, amsmath}

\newcommand{\HRule}[1]{\rule{\linewidth}{#1}} 	

\makeatletter							
\def\printtitle{%
    {\centering \@title\par}}
\makeatother									

\makeatletter							
\def\printauthor{%
    {\centering \large \@author}}				
\makeatother	

\begin{document}

\title{	\normalsize \textsc{Erasmus Mundus Masters in Complex Systems} 	
		 	\\[2.0cm]													
			\HRule{0.5pt} \\										
			\LARGE \textbf{\uppercase{Concentration properties of Gaussian random fields}}	
			\HRule{2pt} \\ [0.5cm]								
			\normalsize July 4, 2012									
		}

\author{
		Girish Sharma\\	
		M2, \'{E}cole Polyechnique\\	
        \texttt{girish.sharma@polytechnique.edu} \\
}

\thispagestyle{empty}				

\printtitle									
  	\vfill
\printauthor								
\vspace{1cm}
\begin{centering}
\emph{Supervisor:} \\
Prof. Philippe \textsc{Mounaix},\\ Centre de Physique Th\'{e}orique \\
\'{E}cole Polytechnique, France\\ 
\end{centering}

\clearpage
\section*{Introduction}
An understanding of stochastic processes has become extremely important to deal with complex phenomena where there is an element of randomness governing the system. Wind waves, random terrain, stock market and turbulence are just a few out of the many areas where an understanding of random phenomena has proved useful to understand these systems. A one dimensional Brownian motion is the simplest form of a stochastic process where we use the interesting properties of the Normal distribution to model stochastic behaviour. For a multivariate Normal distribution the covariance matrix $C_{ij} = \mathbb{E}(X_iX_j)$ defines the properties of the distribution. A Gaussian random field is a generalization of the Brownian motion in more than one dimensions where the covariance matrix $C_{ij}(\bar{x},\bar{y})$ defines the field. There are several texts providing a very good introduction to stochastic processes and Gaussian random fields \cite{Karatzas, Kampen, Adler}.\\

Amplification processes governed by random fields occur in various problems of interest in complex phenomena. A few examples being laser-plasma interaction, reaction-diffusion and turbulent dynamo. In the intermittent regime, the amplified field is dominated by intense fluctuations of a particular functional of the pump field. In order to gain a better understanding of these processes and the underlying physics it is important to characterize these fluctuations. Already some results in this regard have been obtained for a Gaussian scalar field. It is worthwhile to mention a few which are relevant to the problem considered in this report.\\
 
We cite the first example from the domain of lasers which deals with the influence of laser beam on scattering instabilities. Due to the availability of high intensity lasers recently the Hot Spot model \cite{Rose1, Rose2} had been worked out since a perturbative approach breaks down at high intensities. According to this model overall amplification results from amplifications in the small scale high intensity spots but has only been loosely justified. Mounaix and Divol \cite{Divol} proved that overall amplification does not result from successive amplifications as assumed by Hot-Spot model but from a single delocalized mode of the field spreading over the whole interaction length. This is a significant result as it gives a much more accurate description of the gain factor. \\ 

The characterization and realizations of a Gaussian field in the limit when $L^2$-norm is large has also been studied by Mounaix \textit{et al} \cite{Mounaix1}. It has been proved that in the limit of large $L^2$-Norm, the field concentrates onto the eigenspace associated with the largest eigenvalue of the covariance operator of the field. A similar result of Adler states that near a high local maximum the field is governed by its correlation function \cite{Adler}.\\

This project is devoted to generalizing these kinds of results for the case of a Gaussian vector field. We formulate the problem and attempt to solve it in a mathematically rigorous setting deriving a general result and check if we are able to recover back the known results. It is important to note that with this generalization one can further think of some interesting problems one can attempt to solve analytically. As an interesting application of our work we move on to set up the problem of an incompressible turbulent Gaussian flow and examine the structure of the field when the helicity of the flow becomes arbitrarily large.\\

The first chapter deals with all the necessary definitions required and some general notions. We also state three general propositions we proved for a Gaussian vector field. To keep things organized and compact we have provided proofs for the same in Appendix A to C. In the next chapter we recover back Adler's result for a Gaussian field. Then we discuss the application of our results to a homogeneous incompressible Gaussian random flow and calculate the structure of the field when the local helicity of the flow becomes arbitrarily large. Again the details of the calculations are presented in Appendix D.

\clearpage
\section*{Abstract}
\vspace{\fill}
\begin{abstract}
\noindent
We study the problem of a random Gaussian vector field given that a particular real quadratic form $\mathcal{Q}$ is arbitrarily large. We prove that in such a case the Gaussian field is primarily governed by the fundamental eigenmode of a particular operator. As a good check of our proposition we use it to re-derive the result of Adler dealing with the structure of field in the vicinity of a high local maxima. We have also applied our result to an incompressible homogeneous Gaussian random flow in the limit of large local helicity and calculate the structure of the flow.
\end{abstract}
\vspace{\fill}
\clearpage


\tableofcontents 
\clearpage

\section{Definitions and propositions}
\subsection{Complex Gaussian vector field}
In this chapter we introduce necessary definitions and mathematical terminology we will use in the rest of the chapters.  We use the Dirac bra-ket notation. Let $\varphi(x)$ be a zero mean complex vector Gaussian field on a bounded subset $\Lambda$ of $\mathbb{R}^d$. We can write $\varphi(x) \equiv \langle x|\varphi\rangle$ in that notation. In general we have $N\geq 1$ components, so $\varphi_{i}(x)\equiv \langle x,i|\varphi\rangle$. For the covariances, we have Cov$[\varphi_i(x),\varphi_j(y)]=0$ and Cov$[\varphi_i(x),\varphi_j(y)^*]=C_{ij}(x,y)$. Let $f(x)$ be a vector  valued function belonging to $L^2(\Lambda,\{1,...N\})$. The covariance operator $\hat{C}$ acting on $f(x)$ is defined as follows:
\begin{equation}
\langle x,i|\hat{C}|f\rangle = \sum_{j=1}^{N}{\int_{\Lambda}{C_{ij}(x,y)f_j(y)}dy},
\end{equation} 
where $x,y\in\Lambda$. Let $\mathcal{S}\subset L^2(\Lambda,\{1,...N\})$ denote the support of the Gaussian measure of $\varphi(x)$ and $\mathcal{S}^*$ be its dual space. Let $\hat{O}$ be a Hermitian operator acting on $\mathcal{S}^*$ such that for every $\langle\varphi|\in \mathcal{S}^*$ associated with $|\varphi\rangle\in \mathcal{S}$, $\langle\varphi|\hat{O}\in \mathcal{S}^*$. Also for every $|\phi\rangle$ and $|\chi\rangle\in \mathcal{S}$ we have $\langle\phi|\hat{O}|\chi\rangle^* = \langle\chi|\hat{O}|\phi\rangle$. Next we consider the following quadratic form:
\begin{equation}
\mathcal{Q} = \langle\varphi|\hat{O}|\varphi\rangle
\end{equation}
Since $\hat{O}$ is a Hermitian operator, the quadratic form $\mathcal{Q}$ is real.
\subsection{Real Gaussian vector field}
The definitions for real field follow are quite similar to the above. Let $\langle x|\varphi\rangle = \varphi(x)$ be a zero mean real vector Gaussian field on a bounded subset $\Lambda$ of $\mathbb{R}^d$. In general we have $N\geq 1$ components, so $\varphi_{i}(x)\equiv \langle x,i|\varphi\rangle$. For the covariances, we have Cov$[\varphi_i(x),\varphi_j(y)]=C_{ij}(x,y)$. Let $f(x)$ be a vector  valued function belonging to $L^2(\Lambda,\{1,...N\})$. The covariance operator $\hat{C}$ acting on $f(x)$ is defined in the same manner as done previously. Let $\mathcal{S}\subset L^2(\Lambda,\{1,...N\})$ denote the support of the Gaussian measure of $\varphi(x)$ and $\mathcal{S}^*$ be its dual space. Let $\hat{O}$ be a real operator acting on $\mathcal{S}^*$ such that for every $\langle\varphi|\in \mathcal{S}^*$ associated with $|\varphi\rangle\in \mathcal{S}$, $\langle\varphi|\hat{O}\in \mathcal{S}^*$. Also for every $|\phi\rangle$ and $|\chi\rangle\in \mathcal{S}$ we have $\langle\phi|\hat{O}|\chi\rangle^* = \langle\chi|\hat{O}|\phi\rangle$. Next we consider the following quadratic form:
\begin{equation}
\mathcal{Q} = \langle\varphi|\hat{O}|\varphi\rangle \equiv \langle\varphi|\hat{O}^S|\varphi\rangle,
\end{equation}
where $\hat{O}^S = (\hat{O}+\hat{O}^t)/2$ is the symmetric part of $\hat{O}$. Since $\hat{O}^S$ is symmetric, $\mathcal{Q}$ is real.

\subsection{Karhunen-Lo\`{e}ve expansion}
We recall the Karhunen-Lo\`{e}ve expansion for a complex Gaussian field $|\varphi\rangle$ \cite{Adler}.
\begin{equation}
|\varphi\rangle = \sum_{n}{\sqrt{\mu_n}s_n|\mu_n\rangle}
\end{equation}
In the above $\{s_n\}$ is a sequence of i.i.d zero mean complex Gaussian numbers. $\mu_n's$ are the eigenvalues and $|\mu_n\rangle's$ are eigenvectors of the operator $\hat{C}$. For a vector field $|\varphi\rangle$ with $N\geq 1$ components in general the basis vectors $|\mu_n\rangle's$ also have the same number of $N$ components. We can write the above as:
\begin{equation}
|\varphi\rangle = \sum_{n}{s_n\hat{C}^{1/2}|\mu_n\rangle}
\end{equation}
Define the operator $\hat{M} = \hat{C}^{1/2}\hat{O}\hat{C}^{1/2}$. Let $|\lambda_i\rangle$ be the eigenvector of $\hat{M}$ with eigenvalue $\lambda_i$. Since $\hat{M}$ is Hermitian operator $\{|\lambda_i\rangle\}$ form an orthonormal basis. One can write the following relation:
\begin{equation}
|\mu_n\rangle = \sum_{m}{\langle \lambda_m|\mu_n\rangle|\lambda_m\rangle}
\end{equation}
Hence the field $|\varphi\rangle$ can be written as follows:
\begin{equation}
|\varphi\rangle = \sum_m{t_m \hat{C}^{1/2}|\lambda_m\rangle},
\end{equation}
where
\begin{equation}
t_m = \sum_n {s_n\langle\lambda_m|\mu_n\rangle}
\end{equation}
Let us have a look at $\langle t_o^* t_m\rangle$.
\begin{equation}
\begin{split}
\langle t_o^* t_m\rangle &= \sum_n\sum_l \langle \mu_l|\lambda_o\rangle\langle s_l^* s_n\rangle\langle\lambda_m|\mu_n\rangle\\
&= \sum_l {\langle\mu_l|\lambda_o\rangle\langle\lambda_m|\mu_l\rangle}\\
&= \delta_{om}
\end{split}
\end{equation}
The Gaussian field $|\varphi\rangle$ can thus be expressed as:
\begin{equation}
|\varphi\rangle = \sum_i {t_i \hat{C}^{1/2}|\lambda_i\rangle},
\end{equation}
where $t_i$ are i.i.d complex Gaussian random variables with $\langle t_i\rangle = \langle t_i^2\rangle = 0$ and $\langle |t_i|^2\rangle = 1$.  Let us number the positive eigenvalues of $M$ such that $\lambda_1\geq\lambda_2\geq...$ and the negative eigenvalues of $M$ as $\lambda_{-1}\leq\lambda_{-2}\leq ...$. Also let $g_i$ be the degeneracy of $\lambda_i$. We now define 
\begin{equation}
|\bar{\varphi}\rangle = \sum_{i=1}^{g_{\pm 1}}{t_{\pm i}\hat{C}^{1/2}|\lambda_{\pm i}\rangle}
\end{equation}
\begin{equation}
|\delta\varphi\rangle = |\varphi\rangle - |\bar{\varphi}\rangle
\end{equation}
We write $d\mathbb{P}_u$ as the conditional probability measure knowing that $\mathcal{Q} > u$.\\

\subsection{The problem}
We wish to answer the following question in this report: As $\mathcal{Q}\rightarrow\infty$, is there a concentration of $\varphi(x)$ onto some subspace of $\mathcal{S}$ to be determined and, if yes, under what conditions is this true? \\

The answer is given by Propositions 1 and 2 below, according to which:
\begin{equation}
||\delta\varphi||_2^2 / ||\varphi||_2^2\rightarrow 0
\end{equation}
in probability with respect to $d\mathbb{P}_u$ as $u\rightarrow\infty$ (i.e. as $\mathcal{Q}\rightarrow\infty$).
\subsection{Proposition 1} 
Having discussed the necessary definitions in the above sections, we now state the following proposition for a complex vector field $\varphi(x)$. The proof in presented in Appendix A for brevity.\\

\textit{If $\hat{M}$ and $\hat{C}$ are trace class and $||\bar{\varphi}||_2^2 >0$ a.s, then for every $\epsilon>0,$}
\begin{equation}
\lim_{u\rightarrow\infty}\mathbb{\mathbb{P}}_u (||\delta\varphi||_2^2 > \epsilon||\bar{\varphi}||_2^2)=0
\end{equation}
The subscript $u$ in $\mathbb{P}_u$ indicates the probability knowing that $\mathcal{Q}>u$.
\subsection{Proposition 2}
For the real Gaussian field $\varphi(x)$, we discover the same proposition. Again the proof is presented in Appendix B. Note that the proof though follows similar lines as the complex case but is not exactly the same but has a few non-trivial modifications.\\

\textit{If $\hat{M}$ and $\hat{C}$ are trace class and $||\bar{\varphi}||_2^2 >0$ a.s, then for every $\epsilon>0,$}
\begin{equation}
\lim_{u\rightarrow\infty}\mathbb{P}_u (||\delta\varphi||_2^2 > \epsilon||\bar{\varphi}||_2^2)=0
\end{equation}

\subsection{Proposition 3}
Here is a general result we prove dealing with the equivalence of the spectrum of $\hat{M}=\hat{C}^{1/2}\hat{O}\hat{C}^{1/2}$ and $\hat{C}\hat{O}$. We make use the below result in sections 2 and 3. The proof is given in Appendix C.\\

\textit{The spectrum of $\hat{M}$ is equal to the spectrum of the restriction of $\hat{C}\hat{O}$ to $\mathcal{D}(\hat{C}^{-1/2})$.}\\

Here $\mathcal{D}(\hat{C}^{-1/2})$ denotes the domain of $\hat{C}^{-1/2}$.
\clearpage

\section{Application: recovering Adler's result}
We have a random Gaussian field $\varphi(x)$ with correlation function $\hat{C}(x)$. We wish to find what happens to the field when $|\varphi(0)|^2$ is arbitrarily large by applying our proposition. This problem has been attempted by Adler \cite{Adler} who proved that near a high local maxima the field is governed by its correlation function $\hat{C}(x)$. We wish to recover the same result.  As a first step we recall the result of equivalence of the spectra of $\hat{C}\hat{O}$ and $\hat{M} = \hat{C}^{1/2}\hat{O}\hat{C}^{1/2}$ which we stated as proposition 3 in the previous section. For this particular problem, the operator $\hat{O}$ is $|0\rangle\langle 0|$.\\

Next we project the eigenvalue equation for $\hat{C}\hat{O}$ onto $\langle x|$:
\begin{equation}
\langle x|\hat{C}\hat{O}|\phi\rangle = \int{\langle x|\hat{C}\hat{O}|y\rangle \phi(y) dy} = \lambda \phi(x)
\end{equation}
Note that $\langle x|\hat{C}\hat{O}|y\rangle = \langle x|\hat{C}|0\rangle\langle 0|y\rangle = \hat{C}(x)\delta(y)$. Substituting this in the above we obtain:
\begin{equation}
\hat{C}(x)\int{\delta(y)\phi(y) dy} = \lambda \phi(x)
\end{equation}
Hence we obtain: $\lambda\phi(x) = \hat{C}(x)\phi(0)$. If $\phi(0) = 0$ then the eigenvalue $\lambda=0$ or else $\lambda = \hat{C}(0)$ when $\phi(0)$ is different from zero. If $\hat{C}(0) = 1$ (by normalization) then the eigenfunction associated with $\lambda = 1$ is $\phi_1(x) = \phi_1(0)\hat{C}(x)$. We know that $|\phi\rangle = \hat{C}^{1/2}|\lambda\rangle$ (where $|\lambda\rangle$ is the eigenvector of $\hat{M}$ with eigenvalue $\lambda$, see Appendix C). One can write:
\begin{equation}
|\phi_1\rangle = \hat{C}^{1/2}|\lambda_1\rangle = \phi_1(0)\hat{C}|x=0\rangle
\end{equation}
and hence,
\begin{equation}
|\lambda_1\rangle = \phi_1(0)\hat{C}^{1/2}|x=0\rangle
\end{equation}
The eigenvectors of $\hat{M}$ constitute an orthonormal basis, one has the following relation:
\begin{equation}
1 = \langle\lambda_1|\lambda_1\rangle = |\phi_1(0)|^2
\end{equation}
Thus $\phi_1(0)$ is of the form $e^{i\theta}$ - an arbitrary phase, which can be be taken equal to one. Therefore $\phi_1(x) = \hat{C}(x)$ and $\bar{\varphi} = t_1\hat{C}(x)$ and hence $||\bar{\varphi}||_2^2 = t_1^2$.\\

We need to prove that the above is greater that zero (a.s). For this we find out the probability that $\mathbb{P}(|t_1^2|>0)$. This can be done with the standard procedure of writing down the characteristic function and so on:
\begin{equation}
\mathbb{P}(|t_1^2|>0) = \int_{0}^{\infty}{e^{-a}da} = 1
\end{equation}
So $\mathbb{P}(|t_1^2|=0)=0$, therefore $||\bar{\varphi}||_2^2 > 0$ a.s since $\langle 0|\hat{C}^2|0\rangle > 0$.
The field $|\delta\varphi\rangle$ is the following:
\begin{equation}
|\delta\varphi\rangle = |\varphi\rangle - |\bar{\varphi}\rangle
\end{equation}
Since $\hat{M}$ is trace class and $||\bar{\varphi}||_2^2>0$ a.s, we can apply the proposition to obtain the following result for every $\epsilon>0$:
\begin{equation}
\lim_{u\rightarrow\infty}\mathbb{P}_u (||\delta\varphi||_2^2>\epsilon||\bar{\varphi}||_2^2)=0,
\end{equation}
where $d\mathbb{P}_u$ is the conditional probability measure such that $|\varphi(0)|^2>u$.
\clearpage

\section{Application: local helicity of a turbulent fluid}
We consider a stationary homogeneous Gaussian random flow, $\bar{v}(\bar{x})$ with zero mean and correlation function $C_{ij}(\bar{x}-\bar{x}')$. Also we  consider the velocity field $\bar{v}(\bar{x})=\langle x|v\rangle$ on a bounded subset ($\Lambda$) of $\mathbb{R}^3$ with three components i.e. $\bar{v}(\bar{x}) = v_1(\bar{x})\hat{x}_1 + v_2(\bar{x})\hat{x}_2 + v_3(\bar{x})\hat{x}_3$ and $\bar{x} = x_1 \hat{x}_1 +x_2 \hat{x}_2+x_3 \hat{x}_3$. Let $\hat{C}$ the covariance operator governing the Gaussian field be of the following form:
\begin{equation}
C_{ij}(\bar{x}) = \frac{2E}{3}f(x) \delta_{ij} + \frac{E}{3}xf'(x)\left(\delta_{ij} - \frac{x_ix_j}{x^2}\right)
\end{equation}
where $x=|\bar{x}|$, $E$ is the kinetic energy of the turbulent flow per unit mass of the fluid and $f(x)$ is a differentiable even function of $x$ with the following behaviour for small $x$:
\begin{equation}
f(x) = 1-\frac{x^2}{2\lambda^2} + O(x^4)
\end{equation}
where $\lambda$ is "Taylor microscale". Note that we are considering an incompressible, isotropic, and homogeneous turbulent flow which justifies the above form of the correlation function. We use the following definition of local helicity ($h$) at a point $\bar{x}$:
\begin{equation}
h = {\bar{v}(\bar{x})\cdot (\nabla\times \bar{v}(\bar{x}))},
\end{equation}
We are interested in the structure of the flow with large realizations of $|h(0)|$. The above can be expanded as follows:
\begin{equation}
h = {[v_{1}(\partial_{2}v_{3}-\partial_{3}v_{2}) + v_{2}(\partial_{3}v_{1}-\partial_{1}v_{2}) + v_{3}(\partial_{1}v_{2}-\partial_{2}v_{1})]}
\end{equation}
The local helicity at $x=0$ is the following:
\begin{equation}
h(0) = {[v_{1}(0)(\partial_{2}v_{3}-\partial_{3}v_{2})_0 + v_{2}(0)(\partial_{3}v_{1}-\partial_{1}v_{2})_0 + v_{3}(0)(\partial_{1}v_{2}-\partial_{2}v_{1})_0]}
\end{equation}
We have the following definition for the quadratic form $\mathcal{Q}$:
\begin{equation}
\mathcal{Q} = \langle v|\hat{O}|v\rangle,
\end{equation}
where $\hat{O}$ is a symmetric real operator acting on $S^*$, where $S^*$ is the dual of $S\subset L^2(\Lambda,\{1,2,3\})$. Hence the Hilbert space under consideration is a subset of the space of $L^2$ vector functions in $\mathbb{R}^3$. Expanding out, the above can be written as:
\begin{equation}
\mathcal{Q} = \sum_{\mu=1}^3\sum_{\nu=1}^3\int\int{v_{\mu}(\bar{x}) O_{\mu\nu}(\bar{x},\bar{y})v_{\nu}(\bar{y})}d^3x d^3y,
\end{equation}
where $\bar{x}\equiv(x_1,x_2,x_3)\in \mathbb{R}^3$ and $\bar{y}\equiv(y_1,y_2,y_3)\in \mathbb{R}^3$. For further analysis we need to find the appropriate form of the operator $\hat{O}$ admitting the correct form of the local helicity $h(0)$ defined above. Equating the two we find that the following definition of $O_{\mu\nu}$ would suffice:
\begin{equation}
\begin{split}
O_{11}(\bar{x},\bar{y})=0, O_{12}(\bar{x},\bar{y})=-\delta(\bar{x})\delta(\bar{y})\partial_{y_3}, O_{13}(\bar{x},\bar{y})=\delta(\bar{x})\delta(\bar{y})\partial_{y_2}\\
O_{21}(\bar{x},\bar{y})=\delta(\bar{x})\delta(\bar{y})\partial_{y_3}, O_{22}(\bar{x},\bar{y})=0, O_{23}(\bar{x},\bar{y})=-\delta(\bar{x})\delta(\bar{y}))\partial_{y_1}\\
O_{31}(\bar{x},\bar{y})=-\delta(\bar{x})\delta(\bar{y})\partial_{y_2}, O_{32}(\bar{x},\bar{y})=\delta(\bar{x})\delta(\bar{y})\partial_{y_1}, O_{33}(\bar{x},\bar{y})=0
\end{split}
\end{equation}
For instance:
\begin{equation}
\iint{v_1(\bar{x}) O_{12}(\bar{x},\bar{y})v_2(\bar{y})d^3x d^3y} = -\iint{v_1(\bar{x}) \delta(\bar{x})\delta(\bar{y}) \partial_{y_3} v_2(\bar{y}) d^3x d^3y} = -v_1(0)|\partial_{3}v_2|_0
\end{equation}
The above can we written compactly as follows:
\begin{equation}
O_{ij}(\bar{x},\bar{y}) = -\epsilon_{ijk} \delta(\bar{x})\delta(\bar{y})\partial_{y_k},
\end{equation}
where $\epsilon_{ijk}$ is the Levi-Civita symbol. The symmetrized operator $\hat{O}^S$ will be the following:
\begin{equation}
O^S_{ij}(\bar{x},\bar{y}) = -\frac{1}{2}[\epsilon_{ijk} \delta(\bar{x})\delta(\bar{y})\partial_{y_k} + \epsilon_{ijk} \partial_{x_k}\delta(\bar{x})\delta(\bar{y})]
\end{equation}
In the above $\partial_{x_k}$ acts on what is on the left and $\partial_{y_k}$ acts on what is on the right. We wish to write the eigenvalue equation for the operator $\hat{A}=\hat{C}\hat{O}^S$. We have:
\begin{equation}
A_{ik}(\bar{x},\bar{y}) = \sum_j {\int C_{ij}(\bar{x},\bar{z})O^S_{jk}(\bar{z},\bar{y})dz}
\end{equation}
We have the following eigenvalue equation for the operator $\hat{A}$:
\begin{equation}
\langle x|\hat{C}\hat{O}^S|\lambda\rangle = \int {\langle x|\hat{C}\hat{O}^S|y\rangle \lambda(\bar{y}) dy} = \lambda \lambda(\bar{x})
\end{equation}
For the $i^{th}$ component of the field $\lambda(x)$ we have:
\begin{equation}
\sum_{jk}\iint {C_{ik}(\bar{x},\bar{z}) O^S_{kj}(\bar{z},\bar{y}) \lambda_j(\bar{y}) dy dz} = \lambda\lambda_i(\bar{x})
\end{equation}
Substituting for $\hat{O}^S$ we have:
\begin{equation}
\begin{split}
\lambda\lambda_i(\bar{x})
&= -\frac{1}{2}\sum_{jk}\iint {C_{ik}(\bar{x},\bar{z})[\epsilon_{kjl} \delta(\bar{z})\delta(\bar{y})\partial_{y_l} + \epsilon_{kjl} \partial_{z_l}\delta(\bar{z})\delta(\bar{y})]  \lambda_j(\bar{y}) dy dz} \\
&= -\frac{1}{2}\left[\sum_{jk}{C_{ik}(\bar{x},0)\epsilon_{kjl} |\partial_{y_l}\lambda_j(\bar{y})|_0} - \sum_{jk}{\epsilon_{kjl}|\partial_{x_l}C_{ik}(\bar{x},0)|_0 \lambda_j(0)}\right]\\
&= \frac{1}{2} [\bar{c}_i(\bar{x})\cdot (\nabla\times \lambda(0)) - \lambda(0)\cdot (\nabla\times \bar{c}_i(\bar{x}))]
\end{split}
\end{equation}
In the above the $k^{th}$ component of $\bar{c}_i(\bar{x})$ is $C_{ik}(\bar{x})$. Thus we have finally obtained the eigenvalue equation for $\hat{C}\hat{O}$. Next we solve this eigenvalue equation to obtain the results shown below. The details of the calculations are presented in Appendix D. 
The eigenvalue equation can be written as follows after combining all three vector components:
\begin{equation}
\begin{split}
2v\bar{v}(\bar{x}) &= \frac{E}{3}(2f(x) + xf'(x))(\nabla\times \bar{v}(0)) \\&- \frac{E}{3}\bar{x}f'(x)\left(\frac{\bar{x}}{x}\cdot\nabla\times \bar{v}(0)\right) \\&+ \frac{E}{3}\left(4f'(x) + xf''(x)\right)\left(\frac{\bar{x}}{x}\times \bar{v}(0)\right)
\end{split}
\end{equation}
We take the curl of the above to obtain the following (details in Appendix D):
\begin{equation}
\begin{split}
2v (\nabla\times \bar{v}(x)) 
 &=\frac{E}{3}[4f'(x) + xf''(x)] \left(\frac{\bar{x}}{x}\times (\nabla\times \bar{v}(0))\right) \\
 &-\frac{2E}{3x}[4f'(x) + xf''(x)] \bar{v}(0) \\
 &+ \frac{E}{3x}[4f'(x) - 4xf''(x) - x^2f'''(x)]\left[\bar{v}(0) 
  \left(\frac{\bar{x}}{x}\cdot \bar{v}(0)\right)\frac{\bar{x}}{x}\right]
\end{split}
\end{equation}
The above two equations reduce to the following at $\bar{x}=0$ (also by making use of the small ${x}$ behaviour of $f({x})$):
\begin{eqnarray}
v\bar{v}(0) = \frac{E}{3} (\nabla\times \bar{v}(0))\\
\frac{5E}{3\lambda^2} = v (\nabla\times \bar{v}(0))
\end{eqnarray}
Thus,
\begin{equation}
v = \pm\frac{\sqrt{5}E}{3\lambda}
\end{equation}
with,
\begin{equation}
(\nabla\times \bar{v}(0)) = \pm\frac{\sqrt{5}}{\lambda}\bar{v}(0)
\end{equation}
Denoting $\bar{e}_x = \bar{x}/x$ and $\bar{e}_v = \bar{v(0)}/v(0)$ (where $v(0) = |\bar{v}(0)|)$, one obtains the following expression for $\bar{v}(\bar{x})$:
\begin{equation}
\bar{v}(\bar{x}) = \frac{\sqrt{\lambda|h(0)|}}{5^{1/4}}\bar{u}(\bar{x}),
\end{equation}
where,
\begin{equation}
\bar{u}(\bar{x}) = f(x)\bar{e}_v + \frac{x}{2}f'(x)[\bar{e}_v - (\bar{e}_x\cdot\bar{e}_v)\bar{e}_x] \pm \frac{\lambda}{\sqrt{5}}\left[2f'(x) + \frac{x}{2}f''(x)\right] (\bar{e}_x \times \bar{e}_v)
\end{equation}
For $x<<\lambda$ one can substitute the small $x$ approximation for $f(x)$ which gives:
\begin{equation}
\bar{u}(\bar{x}) = \bar{e}_v\pm\frac{\sqrt{5}}{2}\left(\bar{e}_v\times\frac{\bar{x}}{\lambda}\right)-\bar{e}_v\frac{x^2}{\lambda^2} + \frac{(\bar{e}_v\cdot\bar{x})\bar{x}}{2\lambda^2}
\end{equation}
The real Gaussian field $\bar{\varphi}(\bar{x}) = t_1 \bar{v}(\bar{x})$. Now $||\bar{\varphi}||_2^2 = \langle\bar{\varphi}|\bar{\varphi}\rangle = t_1^2||\bar{v}||_2^2$. $||\bar{v}||_2^2$ is the norm of the vector field $\bar{v}(\bar{x})$ which must be positive unless it is a null vector. In the previous section we have proved that $t_1^2>0$ a.s. Hence $||\bar{\varphi}||_2^2 >0$ a.s. We have thus successfully applied the proposition.

\clearpage
\section{Conclusions and perspectives}
In this report we successfully derived a general result for real and complex Gaussian fields and hence were able to answer the question we posed in the beginning i.e what is the structure of the field in the limit of a large quadratic form $\mathcal{Q}$. It is worthwhile to mention that though the proofs for both real and complex fields run on similar lines, we did encounter some special non-trivial issues to deal with in the real case especially the important case when the degeneracy $g_1=1$ (see Appendix). Also note that the case of $g_2=0$ and infinitely degenerate was also treated specially in both cases. It is particularly important because in many relevant physical problems one would encounter this case.\\ 

We conclude that the result we obtained is a generalization to some specific cases treated earlier for example we were successfully able to apply our result to recover a known proposition derived by Adler dealing with the structure of the field in the vicinity of a high local maxima. Though this result seems fairly simple, it is special because it gives an exceedingly good insight to the field structure and a direction to move further.\\

We also state that the problem we dealt with in this report is of relevance in the domain of complex systems and could be employed to generate some important analytical results especially when dealing with random flows. In the last section we already examined an example which dealt with the problem of local helicity of a random flow. We were able to successfully apply our proposition and examine the flow structure. One could extend the above result to try and determine the energy spectrum of the flow $E(k)$. It would require Fourier transforming the velocity field and obtain Fourier modes $u(k)$ from the already obtained $u(x)$.\\

There is also a connection of this problem to the spherical model of a ferromagnet \cite{Berlin} which could be examined and also to Bose-Einstein condensation of a Gaussian vector field \cite{Banerjee} which could be studied further. \\

We believe that this report would certainly provide a very good introduction to someone who wishes to dive more into this field and try and solve more innovative problems !

\clearpage
\appendix
\section{Proof of proposition 1}
\subsection{Proof when $\lambda_{g_1+1}>0$}
We write $d\mathbb{P}_u$ the conditional probability measure knowing that $\mathcal{Q} = \langle\varphi|O|\varphi\rangle > u$. Noting the fact that $\forall a>0$, $P_u(||\delta\varphi||_2^2 > \epsilon||\varphi||_2^2, ||\bar{\varphi}||_2^2 <a) \leq P_u(||\bar{\varphi}||_2^2<a)$, we get the following relation:
\begin{equation}
\mathbb{P}_u(||\delta\varphi||_2^2 > \epsilon||\bar{\varphi}||_2^2) \leq \mathbb{P}_u(||\delta\varphi||_2^2 >\epsilon a) + P_u(||\bar{\varphi}||_2^2<a)
\end{equation}
We denote $\rho(v)$ as the pdf of $\mathcal{Q}_{g_1}=\sum_{n\leq g_1}{\lambda_n |t_n|^2}$. Writing the characteristic function $f(k)$ as $\langle e^{ikQ_{g_1}}\rangle$ we obtain:
\begin{equation}
f(k) = \int {e^{ik\sum_{n\leq g_1}{\lambda_n |t_n|^2}} e^{-\sum_{n\leq g_1}{|t_n|^2}} \prod_{n\leq g_1} d(\Re t_n) d(\Im t_n)}
\end{equation}
The above is a simple Gaussian integral and yields the following result,
\begin{equation}
f(k) = \prod_{n\leq g_1} {\frac{1}{1-ik\lambda_n}}
\end{equation}
leaving aside the normalization factors. Hence we get the expression for $\rho(v)$.
\begin{equation}
\rho(v) = \int {\frac{e^{-ikv}}{(1-ik\lambda_1)^{g_1}}\prod_{n<0}\frac{1}{1-ik\lambda_n}dk}
\end{equation}
For $v>0$ only the pole $k = 1/i\lambda_1$ shall contribute to the integral. The residue being:
\begin{equation}
\begin{split}
\rho(v) &= \frac{1}{(g_1 - 1)!} \lim_{z\rightarrow\frac{1}{i\lambda_1}}\frac{d^{g_1-1}}{dz^{g_1-1}}\frac{e^{-izv}}{\prod_{n<0}(1-iz\lambda_n)}\\
&= \frac{1}{(g_1 - 1)!} \lim_{z\rightarrow\frac{1}{i\lambda_1}}\left(\frac{1}{\prod_{n<0}(1-iz\lambda_n)}\frac{d^{g_1-1}}{dz^{g_1-1}} e^{-izv} \right)+ ...\\
&= \frac{1}{(g_1 - 1)!}\prod_{n<0}\frac{1}{1-\lambda_n/\lambda_1}\left(\frac{v}{\lambda_1}\right)^{g_1-1} \exp(-v/\lambda_1)+ ...
\end{split}
\end{equation}
In the limit of large $v$, only one of the  terms in the derivative of product of two terms survives and it is the one with the large numerical factor of $v^{g_1-1}$. So in the limit $v\rightarrow +\infty$,
\begin{equation}
\rho(v) \sim \frac{1}{(g_1-1)!}\prod_{n<0}\frac{1}{1-\lambda_n/\lambda_1}\left(\frac{v}{\lambda_1}\right)^{g_1-1} \exp(-v/\lambda_1)
\end{equation}
From the above it follows that for every $0<\alpha<1$ there is a $v_0>0$ such that for every $v>v_0$,
\begin{equation}
\rho(v) \geq \frac{1-\alpha}{(g_1-1)!}\prod_{n<0}\frac{1}{1-\lambda_n/\lambda_1}\left(\frac{v}{\lambda_1}\right)^{g_1-1} \exp(-v/\lambda_1)
\end{equation}
From the above we obtain that: 
\begin{equation}
\mathbb{P}(\mathcal{Q}>u)\geq \mathbb{P}(\mathcal{Q}_{g_1}>u) = \int_{u}^{\infty}{\rho(v) dv}
\end{equation}
One needs to evaluate $\int_{u}^{\infty}{v^{g_1}e^{-v/\lambda_1} dv}$. A simple integration by parts yields:
\begin{equation}
\left|\frac{v^{g_1}e^{-v/\lambda_1}}{-\lambda_1}\right|_u^{\infty} - \int{\frac{e^{-v/\lambda_1}}{-\lambda_1}g_1v^{g_1-1}dv}
\end{equation}
Noticing that the subsequent terms are of lesser order than the first one (can be neglected taking the limit $u\rightarrow +\infty$) yields the integral as $(u^{g_1-1}e^{-u/\lambda_1})/\lambda_1$. Hence we obtain the following result:
\begin{equation}
\mathbb{P}(\mathcal{Q}>u)\geq C_1(\alpha) u^{g_1-1} e^{-u/\lambda_1}
\end{equation}
where,
\begin{equation}
C_1(\alpha) = \frac{1-\alpha}{(g_1-1)! \lambda_1}\prod_{n<0}{\frac{1}{1-\lambda_n/\lambda_1}}
\end{equation}
Next we estimate $\mathbb{P}_u(||\bar{\varphi}||_2^2 < a)$. Noting the fact that $\mathbb{P}\left(\sum_i\lambda_i|t_i|^2 >u\right)\leq \mathbb{P}\left(\sum_{i\geq 1}\lambda_i|t_i|^2 >u\right)$, we have:
\begin{equation}
\mathbb{P}_u(||\bar{\varphi}||_2^2 < a)\leq \mathbb{P}\left(\sum_{i,j=1}\langle\lambda_1|C|\lambda_j\rangle t_i^*t_j <a, \sum_{i\geq 1}\lambda_i|t_i|^2 >u\right)
\end{equation}
We know that $\hat{C}$ is the covariance operator and is positive-definite and hermitian (and symmetric in case of real field). Thus the $g_1\times g_1$ matrix $\langle\lambda_i|C|\lambda_j\rangle$ has the same properties. Hence one can rewrite the sum $\sum_{i,j=1}^{g_1}{\langle\lambda_i|c|\lambda_j\rangle t_i^* t_j}$ in the eigen-basis of the $g_1\times g_1$ matrix as:
\begin{equation}
\sum_{i,j=1}^{g_1}{\langle\lambda_i|c|\lambda_j\rangle t_i^* t_j} = \sum_{n=1}^{g_1}{\tilde{\mu}_n}|\bar{t}_n|^2,
\end{equation}
where $\tilde{\mu}_1 \geq \tilde{\mu}_2\geq ...\geq \tilde{\mu}_{g_1}$ are the eigenvalues of the matrix and $\bar{t}_n = \sum_{i=1}^{g_1}\langle\tilde{\mu}_n|\lambda_i\rangle t_i$.
It is easy to see that if $\langle t_i\rangle=\langle t_i^2\rangle = 0$ then $\langle \tilde{t}_i\rangle=\langle \tilde{t}_i^2\rangle = 0$. We have
\begin{equation}
\langle |\tilde{t}_i^2|\rangle = \sum_{n,m=1}^{g_1}\langle\lambda_i|\tilde{\mu}_n\rangle\langle\tilde{\mu}_m|\lambda_i\rangle\langle|t_i|^2\rangle
\end{equation}
Using orthonormality of $\{|\tilde{\mu}_n\rangle's\}$ and $\langle|t_i|^2\rangle = 1$, we have $\langle|\tilde{t}_i|^2\rangle = 1$. Hence $t_i$ and $\tilde{t}_i$ have the same statistical properties. Thus one can drop the tilde in $\tilde{t}_i$. Hence, one obtains:
\begin{equation}
\mathbb{P}(||\bar{\varphi}||_2^2 < a, Q>u) \leq \mathbb{P}\left(\sum_{i=1}^{g_1}{\tilde{\mu}_i|t_i|^2 <a, \sum_{i\geq 1}{\lambda_i |t_i|^2} > u}\right)
\end{equation}
Now $||\bar{\varphi}||_2^2 = \sum_{i=1}^{g_1}{\tilde{\mu}_1|t_i|^2}$. If $||\bar{\varphi}||_2^2>0$ a.s then $\tilde{\mu}_1>0$ a.s since it is the largest eigenvalue of the $g_1\times g_1$ matrix. Since $M$ is assumed to be a trace class so $\sum_{i}|\lambda_i|<+\infty$ a.s, implying that $\lambda_i$ must decrease quickly to zero, hence the degeneracy $g_1$ must be finite. Hence we denote $\tilde{\mu}_{min}$ as the least of $\tilde{\mu}_i > 0$ i.e
\begin{equation}
\tilde{\mu}_{min} = \inf_{1\leq i\leq g_1}\{\tilde{\mu}_i:\tilde{\mu}_i>0\}>0
\end{equation}
Thus we obtain
\begin{equation}
\mathbb{P}(||\bar{\varphi}||_2^2 < a, \mathcal{Q}>u) \leq \mathbb{P}\left(\tilde{\mu}_{min}\sum_{i=1}^{g_1}{|t_i|^2 <a, \sum_{i\geq 1}{\lambda_i |t_i|^2} > u}\right)
\end{equation}
Now,
\begin{eqnarray}
\begin{split}
\mathbb{P}\left(\tilde{\mu}_{min}\sum_{i=1}^{g_1}{|t_i|^2 <a, \sum_{i\geq 1}{\lambda_i |t_i|^2} > u}\right) = \mathbb{P}\left(\sum_{i=1}^{g_1}{|t_i|^2 <\frac{a}{\tilde{\mu}_{min}}, \sum_{i=1}^{g_1}{\lambda_1 |t_i|^2}+\sum_{i> g_1}{\lambda_i |t_i|^2} > u}\right)\\
= \mathbb{P}\left(\sum_{i=1}^{g_1}{|t_i|^2 <\frac{a}{\tilde{\mu}_{min}}, \sum_{i=1}^{g_1}{\lambda_1 |t_i|^2}> u -\sum_{i> g_1}{\lambda_i |t_i|^2} }\right)
\end{split}
\end{eqnarray}
Denoting $\sum_{i> g_1}{\lambda_i |t_i|^2} $ as $x$, we have the above equal to
\begin{equation}
\mathbb{P}\left(\frac{u - x}{\lambda_1} < \sum_{i=1}^{g_1}{|t_i|^2 <\frac{a}{\tilde{\mu}_{min}}}\right)
\end{equation}
Thus one obtains,
\begin{equation}
\mathbb{P}(||\bar{\varphi}||_2^2 < a, \mathcal{Q}>u) \leq  \int_{x=0}^{\infty}{\mathbb{P}\left(\frac{u - x}{\lambda_1}< \sum_{i=1}^{g_1}{|t_i|^2 <\frac{a}{\tilde{\mu}_{min}}}\right) d\mathbb{P}\left( \sum_{i> g_1}{\lambda_i |t_i|^2} = x\right)}
\end{equation}
Now if $x<u-\frac{\lambda_1 a}{\tilde{\mu}_{min}}$ then $\frac{u-x}{\lambda_1} > \frac{a}{\tilde{\mu}_{min}}$ which is not possible. Thus the above can be further simplified to:
\begin{equation}
\begin{split}
\int_{x=u-\frac{\lambda_1 a}{\tilde{\mu}_{\min}}}^{\infty}{\mathbb{P}\left(\frac{u - x}{\lambda_1}< \sum_{i=1}^{g_1}{|t_i|^2 <\frac{a}{\tilde{\mu}_{min}}}\right) d\mathbb{P}\left( \sum_{i> g_1}{\lambda_i |t_i|^2} = x\right)}\leq \int_{x=u-\frac{\lambda_1 a}{\tilde{\mu}_{\min}}}^{\infty} {d\mathbb{P}\left( \sum_{i> g_1}{\lambda_i |t_i|^2} = x\right)}\\
=\mathbb{P}\left(\sum_{i>g_1}\lambda_i |t_i|^2 > u - \frac{\lambda_1 a}{\tilde{\mu}_{min}}\right)
\end{split}
\end{equation}
We recall the exponential Markov inequality that for a random variable X, for every $t>0$, $\mathbb{P}(X>a)\leq e^{-ta}\mathbb{E}[e^{tX}]$. Applying it to the above we obtain $\forall c>0$:
\begin{equation}
\mathbb{P}\left(\sum_{i>g_1}\lambda_i |t_i|^2 > u - \frac{\lambda_1 a}{\tilde{\mu}_{min}}\right) \leq e^{-c(u-\lambda_1a/\tilde{\mu}_{min})} \mathbb{E}\left[\exp\left( c \sum_{i>g_1}\lambda_i |t_i|^2\right)\right]
\end{equation}
Now,
\begin{equation}
\mathbb{E}\left[\exp\left( c \sum_{i>g_1}\lambda_i |t_i|^2\right)\right] = \mathbb{E}\left[\prod_{i>g_1}\exp\left(c\lambda_i |t_i|^2\right)\right]
\end{equation}
Therefore,
\begin{equation}
\begin{split}
\mathbb{E}\left[\exp\left(c\lambda_i |t_i|^2\right)\right] = \iint e^{c\lambda_i|t_i|^2}d\mathbb{P}(\Re t_i)d\mathbb{P}(\Im t_i) \\
= \iint e^{c\lambda_it_{iR}^2} e^{c\lambda_i t_{iI}^2} e^{-t_{iR}^2} e^{-t_{iI}^2} \frac{dt_{iR} dt_{iI}}{\pi} = \frac{1}{1-c\lambda_i}
\end{split}
\end{equation}
apart from some normalization constants. Hence,
\begin{equation}
\mathbb{P}\left(\sum_{i>g_1}\lambda_i |t_i|^2 > u - \frac{\lambda_1 a}{\tilde{\mu}_{min}}\right) \leq e^{-c(u-\lambda_1a/\tilde{\mu}_{min})} \prod_{i>g_1} \frac{1}{1-c\lambda_i}
\end{equation}
Also note that the above product is necessarily well behaved if $c<1/\lambda_{g_1+1}$, hence we impose this constraint on $c$. Now we can obtain an upper bound on $P_u(||\bar{\varphi}||_2^2 < a)\equiv \frac{P(||\bar{\varphi}||_2^2 < a, Q>u)}{P(Q>u)}$ since we already have a bound on $P(Q>u)$. We take $c = (\lambda_1^{-1} + \lambda_{g_1+1}^{-1})/2$ and substitute in the above equation.
\begin{equation}
\exp\left[-c(u-\lambda_1a/\tilde{\mu}_{min})\right] = \exp\left[-\left(\frac{1}{\lambda_1} + \frac{1}{\lambda_{g_1+1}}\right)\frac{u}{2}\right]\exp\left[\frac{a(\lambda_1 + \lambda_{g_1+1})}{2\lambda_{g_1+1}\tilde{\mu}_{min}}\right]
\end{equation}
Thus,
\begin{equation}
\mathbb{P}_{u}((||\bar{\varphi}||_2^2 < a) \leq C_2(\alpha)\exp\left[-\left(\frac{1}{\lambda_1} + \frac{1}{\lambda_{g_1+1}}\right)\frac{u}{2}\right],
\end{equation}
where,
\begin{equation}
C_2(\alpha) = \exp\left[\frac{a(\lambda_1 + \lambda_{g_1+1})}{2\lambda_{g_1+1}\tilde{\mu}_{min}}\right]\prod_{i>g_1}\frac{1}{1-(\lambda_1^{-1} + \lambda_{g_1+1}^{-1})\lambda_i/2}
\end{equation}
Using the bound on $\mathbb{P}(\mathcal{Q}>u)$, we finally obtain:
\begin{equation}
\mathbb{P}_u(||\bar{\varphi}||_2^2 < a)\leq \frac{C_2(\alpha)}{C_1(\alpha)}\exp\left[-\left(\frac{1}{\lambda_{g_1+1}} - \frac{1}{\lambda_1}\right)\frac{u}{2}\right]
\end{equation}
which equals zero in the limit $u\rightarrow\infty$.\\ 

Next we estimate the conditional probability $\mathbb{P}_u(||\delta\varphi||_2^2>\epsilon a)$. Using the fact that $P(Q>u)\leq P\left(\sum_{i\geq 1}{\lambda_i |t_i|^2}>u\right)$, we write the joint probability measure of $t_{i\notin [1,g_1]}$ as:
\begin{equation}
d\mathbb{P}\left(t_{i\notin [1,g_1]},\sum_{i}{\lambda_i|t_i|^2 >u}\right)\leq d\mathbb{P}\left(t_{i\notin [1,g_1]},\sum_{i\geq 1}{\lambda_i|t_i|^2 >u}\right)
\end{equation}
Also,
\begin{equation}
\mathbb{P}\left(\sum_{i\geq 1}{\lambda_i|t_i|^2 >u}\right) = \mathbb{P} \left(\lambda_1\sum_{i=1}^{g_1}{|t_i|^2}>u-\sum_{i>g_1}{\lambda_i|t_i|^2}\right),
\end{equation}
and thus we can write the following relation:
\begin{equation}
d\mathbb{P}\left(t_{i\notin [1,g_1]},\sum_{i}{\lambda_i|t_i|^2 >u}\right)\leq \mathbb{P} \left(\sum_{i=1}^{g_1}{|t_i|^2}>\frac{u}{\lambda_1}-\sum_{i>g_1}{\frac{\lambda_i}{\lambda_1}|t_i|^2}\right)\prod_{i\notin[1,g_1]}e^{-|t_i|^2}\frac{d^2t_i}{\pi}
\end{equation}
Evaluation of $d\mathbb{P}\left(\sum_{i=1}^{g_1}{|t_i|^2}\right)$ follows the same procedure as we followed earlier - writing the characteristic function and taking the inverse Fourier transform. One obtains a $g_1$ order pole at $k=-i$ and we obtain the following pdf:
\begin{equation}
\rho(v) = \int {\frac{H(v) v^{g_1-1}}{(g_1-1)!}e^{-v}dv}
\end{equation}
where $H(v)$ is the Heavyside step function. Hence,
\begin{equation}
\mathbb{P} \left(\sum_{i=1}^{g_1}{|t_i|^2}>\frac{u}{\lambda_1}-\sum_{i>g_1}{\frac{\lambda_i}{\lambda_1}|t_i|^2}\right) = \int_{\frac{u}{\lambda_1}-\sum_{i>g_1}{\frac{\lambda_i}{\lambda_1}|t_i|^2}}^{\infty}{\frac{H(v) v^{g_1-1}}{(g_1-1)!}e^{-v}dv}
\end{equation}
Performing a change of variables $v\rightarrow v-\sum_{i>g_1}{\lambda_i|t_i|^2/\lambda_1}$ and using the fact that $H(v)v^{g_1-1}$ is an increasing function of $v$ we obtain:
\begin{equation}
\mathbb{P} \left(\sum_{i=1}^{g_1}{|t_i|^2}>\frac{u}{\lambda_1}-\sum_{i>g_1}{\frac{\lambda_i}{\lambda_1}|t_i|^2}\right)\leq\frac{1}{(g_1-1)!}\int_{u/\lambda_1}^{\infty}{v^{g_1-1}e^{-v}dv \prod_{i>g_1}e^{\lambda_i|t_i|^2/\lambda_1}}
\end{equation}
We perform integration by parts and note that the subsequent terms are of lesser order than the first one (can be neglected taking the limit $u\rightarrow +\infty$) and obtain:
\begin{equation}
\int_{u/\lambda_1}^{\infty}{v^{g_1-1}e^{-v}dv}\sim \left(\frac{u}{\lambda_1}\right)^{g_1-1}e^{-u/\lambda_1}
\end{equation}
Thus for every $\alpha>0$, there is a $v_1>0$ such that for every $u>v_1$,
\begin{equation}
\mathbb{P} \left(\sum_{i=1}^{g_1}{|t_i|^2}>\frac{u}{\lambda_1}-\sum_{i>g_1}{\frac{\lambda_i}{\lambda_1}|t_i|^2}\right)\leq\frac{1+\alpha}{(g_1-1)!}{\left(\frac{u}{\lambda_1}\right)^{g_1-1}e^{-u/\lambda_1} \prod_{i>g_1}e^{\lambda_i|t_i|^2/\lambda_1}}
\end{equation}
Thus for $u$ very large we have:
\begin{equation}
d\mathbb{P}\left(t_{i\notin [1,g_1]},\sum_{i}{\lambda_i|t_i|^2 >u}\right)\leq \frac{1+\alpha}{(g_1-1)!}{\left(\frac{u}{\lambda_1}\right)^{g_1-1}e^{-u/\lambda_1} \prod_{i>g_1}e^{\lambda_i|t_i|^2/\lambda_1}} \prod_{i\notin[1,g_1]}e^{-|t_i|^2}d^2t_i
\end{equation}
Making use of the bound on $\mathbb{P}_u(\mathcal{Q}>u)$, we can write the conditional probability measure as:
\begin{equation}
d\mathbb{P}_u\left(t_{i\notin[1,g_1]}\right) \leq C_3(\alpha) \left(\prod_{i<0}{e^{-|t_i|^2}\frac{d^2t_i}{\pi}}\right)\left(\prod_{i>g_1}{e^{-(1-\lambda_i/\lambda_1)|t_i|^2}\frac{d^2t_i}{\pi}}\right)
\end{equation}
where,
\begin{equation}
C_3(\alpha) = \frac{(1+\alpha)}{(1-\alpha)}\lambda_1^{-g_1}\prod_{n<0}{(1-\lambda_n/\lambda_1)}
\end{equation}
Next we estimate $\mathbb{E}_{u}[e^{c||\delta\varphi||_2^2}]$ for some $c>0$.
\begin{equation}
\mathbb{E}_{u}[e^{c||\delta\varphi||_2^2}] \leq C_3(\alpha) \int{e^{c\sum_{i,j\notin[1,g_1]}{t^*_i\langle\lambda_i|C|\lambda_j\rangle}t_j}}\left(\prod_{i<0}{e^{-|t_i|^2}\frac{d^2t_i}{\pi}}\right)\left(\prod_{i>g_1}{e^{-(1-\lambda_i/\lambda_1)|t_i|^2}\frac{d^2t_i}{\pi}}\right)
\end{equation}
We can write the above in a compact form as follows:
\begin{equation}
\mathbb{E}_{u}[e^{c||\delta\varphi||_2^2}] \leq C_3(\alpha) \int{e^{-\sum_{i,j\notin[1,g_1]}{t^*_iG_{ij}t_j}}\prod_{i,j\notin[1,g_1]}\frac{d^2t_i}{\pi}}
\end{equation}
where $\hat{G}$ is the matrix below
\begin{equation}
G_{ij} = diag(\min\{1,1-\lambda_i/\lambda_1\}) - c\langle\lambda_i|\hat{C}|\lambda_j\rangle
\end{equation}
For the above Gaussian integral to exist, the matrix $\hat{G}$ must be strictly positive-definite. When we perform the Gaussian integration we end up with an infinite determinant i.e. an infinite product of the eigenvalues which needs to be finite. Thus the matrix diag $(max\{0,\lambda_i/\lambda_1\}) + c\langle\lambda_i|\hat{C}|\lambda_j\rangle$ needs to be a trace class. From which it follows that $\hat{M}$ and $\hat{C}$ must be trace class operators. Noticing that $||\hat{C}|| = \mu_1$ and $\min\{1,1-\lambda_i/\lambda_1\}\geq1-\lambda_{g_1+1}/\lambda_1$ for $i\notin[1,g_1]$, then this condition is satisfied for $c<(1-\lambda_{g1+1}/\lambda_1)/\mu_1$. Thus there exists a $c>0$ such that the Gaussian integral exists-which we call as $C_4(c)$. Thus,
\begin{equation}
\mathbb{E}_{u}[e^{c||\delta\varphi||_2^2}] \leq C_3(\alpha) C_4(c)
\end{equation}
Using exponential Markov inequality, we obtain:
\begin{equation}
\mathbb{P}_u(||\delta\varphi||_2^2>\epsilon a)\leq e^{-\epsilon ca}E_{u}[e^{c||\delta\varphi||_2^2}]\leq e^{-\epsilon ca} C_3(\alpha) C_4(c)
\end{equation}
Thus plugging both estimates into our original equation and taking the limit $u\rightarrow\infty$ we obtain:
\begin{equation}
\lim_{u\rightarrow\infty}\mathbb{P}_u(||\delta\varphi||_2^2>\epsilon||\varphi||_2^2)\leq e^{-\epsilon ca} C_3(\alpha) C_4(c)
\end{equation}
In the above $a$ can be arbitrarily large and leads to the conclusion that $\lim_{u\rightarrow\infty}P_u(||\delta\varphi||_2^2>\epsilon||\varphi||_2^2) = 0$.
\subsection{Proof for $\lambda_{g_1+1}=0$}
There is no modification until we obtain the following lower bound on $\mathbb{P}(\mathcal{Q}>u)$:
\begin{equation}
\mathbb{P}(\mathcal{Q}>u)\geq C_1(\alpha) u^{g_1-1}\exp\left(\frac{-u}{\lambda_1}\right)
\end{equation}
Next we estimate the conditional probability $\mathbb{P}_u(||\bar{\varphi}||_2^2<a)$ and obtain the following relation:
\begin{equation}
\mathbb{P}(||\bar{\varphi}||_2^2<a,\mathcal{Q}>u)\leq \mathbb{P}\left(\bar{\mu}_{min}\sum_{i=1}^{g_1}{|t_i|^2}<a, \sum_{i\geq 1}{\lambda_i|t_i|^2}>u\right)
\end{equation}
Since $\lambda_{g_1+1}$ is zero and infinitely degenerate, we have:
\begin{equation}
\mathbb{P}\left(\bar{\mu}_{min}\sum_{i=1}^{g_1}{|t_i|^2}<a, \sum_{i\geq 1}{\lambda_i|t_i|^2}>u\right) = \mathbb{P}\left(\bar{\mu}_{min}\sum_{i=1}^{g_1}{|t_i|^2}<a, \sum_{i=1}^{g_1}{\lambda_i|t_i|^2}>u\right)
\end{equation}
Thus we need to evaluate the following:
\begin{equation}
\mathbb{P}\left(\frac{u}{\lambda_1} < \sum_{i=1}^{g_1}{|t_i|^2} < \frac{a}{\bar{\mu}_{min}}\right) 
\end{equation}
We note that the above is zero for $u>\lambda_1 a/{\bar{\mu}_{min}}$. So $\mathbb{P}(||\bar{\varphi}||_2^2<a, \mathcal{Q}>u) = 0$ for $u>\lambda_1 a/{\bar{\mu}_{min}}$. Hence we have the result that for every $a>0$, for $u>\lambda_1 a/{\bar{\mu}_{min}}$, $\mathbb{P}_u(||\bar{\varphi}||_2^2<a) = 0$.\\

We next move on to the part where we estimate $\mathbb{P}_{u}(||\delta\varphi||_2^2 > \epsilon a)$. We write the following conditional probability measure:
\begin{equation}
d\mathbb{P}\left(t_{i\neq [1,g_1]},\mathcal{Q}>u\right)\leq d\mathbb{P}\left(t_{i\neq [1,g_1]},\sum_{i\geq 1}{\lambda_i |t_i|^2}>u\right)
\end{equation}
Since $\lambda_{g_1+1}$ is zero and infinitely degenerate, we have:
\begin{equation}
d\mathbb{P}\left(t_{i\neq [1,g_1]},\sum_{i\geq 1}{\lambda_i |t_i|^2}>u\right) = d\mathbb{P}\left(t_{i\neq [1,g_1]},\sum_{i=1}^{g_1}{\lambda_i |t_i|^2}>u\right) = \mathbb{P}\left(\sum_{i=1}^{g_1}{|t_i|^2}>\frac{u}{\lambda_1}\right)\prod_{i\neq [1,g_1]}{\frac{e^{-|t_i|^2}}{\pi}d^2t_i}
\end{equation}
For large $u$ one has (also since $H(v)=1$ in the considered domain (see below)):
\begin{equation}
\mathbb{P}\left(\sum_{i=1}^{g_1}{|t_i|^2}>\frac{u}{\lambda_1}\right) = \int_{u/\lambda_1}^{\infty}{\frac{H(v) v^{g_1-1}}{(g_1-1)!}e^{-v}dv}\sim\frac{1}{(g_1-1)!}\left(\frac{u}{\lambda_1}\right)^{g_1-1}e^{-u/\lambda_1}
\end{equation}
It follows that for every $\alpha>0$, there is a $v_1>0$ such that for every $v>v_1$,
\begin{equation}
\mathbb{P}(||\bar{\varphi}||_2^2<a,\mathcal{Q}>u)\leq \frac{1+\alpha}{(g_1-1)!}\left(\frac{u}{\lambda_1}\right)^{g_1-1}e^{-u/\lambda_1}\prod_{i\neq [1,g_1]}{\frac{e^{-|t_i|^2}}{\pi}d^2t_i}
\end{equation}
The conditional probability (for $u>\max\{v_0,v_1\}$) will be:
\begin{equation}
d\mathbb{P}_u (\{t_{i\neq [1,g_1]}\})\leq C_3(\alpha)\prod_{i\neq [1,g_1]}{\frac{e^{-|t_i|^2}}{\pi}d^2t_i},
\end{equation}
where $C_3(\alpha)$ is the same as defined in the case of $\lambda_{g_1+1}> 0$. Thus one obtains for large $u$, the following: 
\begin{equation}
\mathbb{E}_u[\exp(c||\delta\varphi||_2^2)]\leq C_3(\alpha)C_4(c),
\end{equation}
where $C_4(c)$ is a Gaussian integral whose existence is ensured if we choose $c<1/\mu_1$ where $||\hat{C}||=\mu_1$ and $\hat{C}$ being trace class.\\

\clearpage
\section{Proof of proposition 2}
\subsection{Proof when $\lambda_{g_1+1}>0$}
We write $d\mathbb{P}_u$ the conditional probability measure knowing that $\mathcal{Q} = \langle\varphi|O|\varphi\rangle > u$. Using the fact that $\forall a>0$, $P_u(||\delta\varphi||_2^2 > \epsilon||\varphi||_2^2, ||\bar{\varphi}||_2^2 <a) \leq P_u(||\bar{\varphi}||_2^2<a)$, we get the following relation:
\begin{equation}
\mathbb{P}_u(||\delta\varphi||_2^2 > \epsilon||\bar{\varphi}||_2^2) \leq \mathbb{P}_u(||\delta\varphi||_2^2 >\epsilon a) + \mathbb{P}_u(||\bar{\varphi}||_2^2<a)
\end{equation}
The quadratic form $\mathcal{Q}=\langle\varphi|\hat{O}^S|\varphi\rangle$ can be written as:
\begin{equation}
\mathcal{Q} = \sum_{i}{t^2_i \lambda_i}
\end{equation}
Following the proof of the complex case we evaluate the pdf ($\rho(v)$)of $\mathcal{Q}_{g_1}=\sum_{n<g_1}{\lambda_n t^2_n}$. Writing the characteristic  function $f(k)$ as $\langle e^{ik\mathcal{Q}_{g_1}}\rangle$ we obtain:
\begin{equation}
f(k) = \int {e^{ik\sum_{n\leq g_1}{\lambda_n t_n^2}} e^{-\sum_{n\leq g_1}{t_n^2}} \prod_{n\leq g_1} dt_n}
\end{equation}
This can be evaluated to be:
\begin{equation}
f(k) = \prod_{n\leq g_1} {\frac{1}{\sqrt{1-ik\lambda_n}}}
\end{equation}
leaving aside the normalization factors. Hence we get the expression for $\rho(v)$.
\begin{equation}
\rho(v) = \int_{-\infty}^{\infty} {\frac{e^{-ikv}}{(\sqrt{1-2ik\lambda_1})^{g_1}}\prod_{n<0}\frac{1}{\sqrt{1-2ik\lambda_n}}\frac{dk}{2\pi}}
\end{equation}
However in the above integral we note the fact that for large $v$ only the vicinity of $-i/2\lambda_1$ contributes. In that limit one can write:
\begin{equation}
\rho(v) \sim \prod_{n<0}{\frac{1}{\sqrt{1-\lambda_n/\lambda_1}}} \int_{-\infty}^{\infty}{\frac{e^{-ikv}}{(1-2ik\lambda_1)^{g_1/2}}\frac{dk}{2\pi}} 
\end{equation}
Using the change of variables $s=1-2ik\lambda_1$, we have:
\begin{equation}
\rho(v) \sim \frac{e^{-v/2\lambda_1}}{2\lambda_1}\prod_{n<0}\frac{1}{\sqrt{1-\lambda_n/\lambda_1}}\int_{1-i\infty}^{1+i\infty}{\frac{e^{vs/2\lambda_1}}{s^{g_1/2}}\frac{ds}{2i\pi}}
\end{equation}
We note that $M$ being a trace class ensures that the infinite product before the integral exists. The integral can be solved to obtain the following \cite{Abramowitz}:
\begin{equation}
\rho(v) \sim \frac{1}{(g_1/2-1)!}\prod_{n<0}\frac{1}{\sqrt{1-\lambda_n/\lambda_1}}\left(\frac{v}{2\lambda1}\right)^{g_1/2-1}\frac{e^{-v/2\lambda_1}}{2\lambda_1}
\end{equation}
From the above it follows that for every $0<\alpha<1$ there is a $v_0>0$ such that for every $v>v_0$,
\begin{equation}
\rho(v) \geq \frac{(1-\alpha)}{(g_1/2-1)!}\prod_{n<0}\frac{1}{\sqrt{1-\lambda_n/\lambda_1}}\left(\frac{v}{2\lambda1}\right)^{g_1/2-1}\frac{e^{-v/2\lambda_1}}{2\lambda_1}
\end{equation}
From the above we obtain that: 
\begin{equation}
\mathbb{P}(\mathcal{Q}>u)\geq \mathbb{P}(\mathcal{Q}_{g_1}>u) = \int_{u}^{\infty}{\rho(v) dv}
\end{equation}
One needs to evaluate $\int_{u}^{\infty}{v^{\frac{g_1}{2}-1}e^{-v/2\lambda_1} \frac{dv}{2\lambda_1}}$. A simple integration by parts and noticing that the subsequent terms are of lesser order than the first one (can be neglected taking the limit $u\rightarrow +\infty$) yields the integral as $(u^{\frac{g_1}{2}-1}e^{-u/2\lambda_1})$. Thus from the asymptotics one has:
\begin{equation}
\int_{u}^{\infty} {v^{g_1/2-1}e^{-v/2\lambda_1}\frac{dv}{2\lambda_1}} \sim u^{g_1/2-1}e^{-u/2\lambda_1}
\end{equation}
From the above it follows that there is a $v_1>0$ such that for every $u>v_1$,
\begin{equation}
\int_{u}^{\infty} {v^{g_1/2-1}e^{-v/2\lambda_1}}\frac{dv}{2\lambda_1}\geq (1-\alpha) u^{g_1/2-1}e^{-u/2\lambda_1}
\end{equation}
From the above results one obtains for $u>$ max$\{v_0,v_1\}$,
\begin{equation}
\mathbb{P}(\mathcal{Q}>u) \geq C_1(\alpha) u^{g_1/2-1}e^{-u/2\lambda_1},
\end{equation}
where
\begin{equation}
C_1(\alpha) = \frac{(1-\alpha)^2}{(g_1/2-1)! (2\lambda_1)^{g_1/2-1}}\prod_{n<0}\frac{1}{1-\lambda_n/\lambda_1}
\end{equation}
Next we estimate $\mathbb{P}_u(||\bar{\varphi}||_2^2 < a)$. Noting the fact that $\mathbb{P}\left(\sum_i\lambda_it_i^2 >u\right)\leq \mathbb{P}\left(\sum_{i\geq 1}\lambda_it_i^2 >u\right)$, we have:
\begin{equation}
\mathbb{P}_u(||\bar{\varphi}||_2^2 < a)\leq \mathbb{P}\left(\sum_{i,j=1}\langle\lambda_1|C|\lambda_j\rangle t_it_j <a, \sum_{i\geq 1}\lambda_it_i^2 >u\right)
\end{equation}
We know that $\hat{C}$ is the covariance operator and is positive-definite and symmetric in case of real field. Thus the $g_1\times g_1$ matrix $\langle\lambda_i|C|\lambda_j\rangle$ has the same properties. Hence one can rewrite the sum $\sum_{i,j=1}^{g_1}{\langle\lambda_i|c|\lambda_j\rangle t_i t_j}$ in the eigen-basis of the $g_1\times g_1$ matrix as:
\begin{equation}
\sum_{i,j=1}^{g_1}{\langle\lambda_i|c|\lambda_j\rangle t_i t_j} = \sum_{n=1}^{g_1}{\tilde{\mu}_n}\bar{t}_n^2,
\end{equation}
where $\tilde{\mu}_1 \geq \tilde{\mu}_2\geq ...\geq \tilde{\mu}_{g_1}$ are the eigenvalues of the matrix and $\bar{t}_n = \sum_{i=1}^{g_1}\langle\tilde{\mu}_n|\lambda_i\rangle t_i$.
It is easy to see that if $\langle t_i\rangle = 0$ then $\langle \tilde{t}_i\rangle= 0$ by simple substitution. We have
\begin{equation}
\langle \tilde{t}_i^2\rangle = \sum_{n,m=1}^{g_1}\langle\lambda_i|\tilde{\mu}_n\rangle\langle\tilde{\mu}_m|\lambda_i\rangle\langle t_i^2\rangle
\end{equation}
Using orthonormality of $\{|\tilde{\mu}_n\rangle's\}$ and $\langle t_i^2\rangle = 1$, we have $\langle\tilde{t}_i^2\rangle = 1$. Hence $t_i$ and $\tilde{t}_i$ have the same statistical properties. Thus one can drop the tilde in $\tilde{t}_i$. Hence, one obtains:
\begin{equation}
\mathbb{P}(||\bar{\varphi}||_2^2 < a, \mathcal{Q}>u) \leq \mathbb{P}\left(\sum_{i=1}^{g_1}{\tilde{\mu}_it_i^2 <a, \sum_{i\geq 1}{\lambda_i t_i^2} > u}\right)
\end{equation}
Now $||\bar{\varphi}||_2^2 = \sum_{i=1}^{g_1}{\tilde{\mu}_1t_i^2}$. If $||\bar{\varphi}||_2^2>0$ a.s then $\tilde{\mu}_1>0$ a.s since it is the largest eigenvalue of the $g_1\times g_1$ matrix. Since $M$ is assumed to be a trace class so $\sum_{i}|\lambda_i|<+\infty$ a.s, implying that $\lambda_i$ must decrease quickly to zero, hence the degeneracy $g_1$ must be finite. Hence we denote $\tilde{\mu}_{min}$ as the least of $\tilde{\mu}_i > 0$ i.e
\begin{equation}
\tilde{\mu}_{min} = \inf_{1\leq i\leq g_1}\{\tilde{\mu}_i:\tilde{\mu}_i>0\}>0
\end{equation}
Thus we obtain
\begin{equation}
\mathbb{P}(||\bar{\varphi}||_2^2 < a, \mathcal{Q}>u) \leq \mathbb{P}\left(\tilde{\mu}_{min}\sum_{i=1}^{g_1}{t_i^2 <a, \sum_{i\geq 1}{\lambda_i t_i^2} > u}\right)
\end{equation}
Now,
\begin{eqnarray}
\begin{split}
\mathbb{P}\left(\tilde{\mu}_{min}\sum_{i=1}^{g_1}{t_i^2 <a, \sum_{i\geq 1}{\lambda_i t_i^2} > u}\right) = \mathbb{P}\left(\sum_{i=1}^{g_1}{t_i^2 <\frac{a}{\tilde{\mu}_{min}}, \sum_{i=1}^{g_1}{\lambda_1 t_i^2}+\sum_{i> g_1}{\lambda_i t_i^2} > u}\right)\\
= \mathbb{P}\left(\sum_{i=1}^{g_1}{t_i^2 <\frac{a}{\tilde{\mu}_{min}}, \sum_{i=1}^{g_1}{\lambda_1 t_i^2}> u -\sum_{i> g_1}{\lambda_i t_i^2} }\right)
\end{split}
\end{eqnarray}
Denoting $\sum_{i> g_1}{\lambda_i t_i^2} $ as $x$, we have the above equal to
\begin{equation}
\mathbb{P}\left(\frac{u - x}{\lambda_1} < \sum_{i=1}^{g_1}{t_i^2 <\frac{a}{\tilde{\mu}_{min}}}\right)
\end{equation}
Thus one obtains,
\begin{equation}
\mathbb{P}(||\bar{\varphi}||_2^2 < a, \mathcal{Q}>u) \leq  \int_{x=0}^{\infty}{\mathbb{P}\left(\frac{u - x}{\lambda_1}< \sum_{i=1}^{g_1}{t_i^2 <\frac{a}{\tilde{\mu}_{min}}}\right) d\mathbb{P}\left( \sum_{i> g_1}{\lambda_i t_i^2} = x\right)}
\end{equation}
Now if $x<u-\frac{\lambda_1 a}{\tilde{\mu}_{min}}$ then $\frac{u-x}{\lambda_1} > \frac{a}{\tilde{\mu}_{min}}$ which is not possible. Thus the above can be further simplified to:
\begin{equation}
\begin{split}
\int_{x=u-\frac{\lambda_1 a}{\tilde{\mu}_{\min}}}^{\infty}{\mathbb{P}\left(\frac{u - x}{\lambda_1}< \sum_{i=1}^{g_1}{t_i^2 <\frac{a}{\tilde{\mu}_{min}}}\right) d\mathbb{P}\left( \sum_{i> g_1}{\lambda_i t_i^2} = x\right)}\leq \int_{x=u-\frac{\lambda_1 a}{\tilde{\mu}_{\min}}}^{\infty} {d\mathbb{P}\left( \sum_{i> g_1}{\lambda_i t_i^2} = x\right)}\\
=\mathbb{P}\left(\sum_{i>g_1}\lambda_i t_i^2 > u - \frac{\lambda_1 a}{\tilde{\mu}_{min}}\right)
\end{split}
\end{equation}
We recall the exponential Markov inequality that for a random variable $X$, for every $t>0$, $\mathbb{P}(X>a)\leq e^{-ta}\mathbb{E}[e^{tX}]$. Applying it to the above we obtain $\forall c>0$:
\begin{equation}
\mathbb{P}\left(\sum_{i>g_1}\lambda_i t_i^2 > u - \frac{\lambda_1 a}{\tilde{\mu}_{min}}\right) \leq e^{-c(u-\lambda_1a/\tilde{\mu}_{min})} \mathbb{E}\left[\exp\left( c \sum_{i>g_1}\lambda_i t_i^2\right)\right]
\end{equation}
Now,
\begin{equation}
\mathbb{E}\left[\exp\left( c \sum_{i>g_1}\lambda_i t_i^2\right)\right] = \mathbb{E}\left[\prod_{i>g_1}\exp\left(c\lambda_i t_i^2\right)\right]
\end{equation}
Therefore,
\begin{equation}
\begin{split}
\mathbb{E}\left[\exp\left(c\lambda_i t_i^2\right)\right] = \int e^{c\lambda_it_i^2}d\mathbb{P}(t_i) \\
= \int e^{c\lambda_it_{i}^2}  e^{-t_{i}^2/2}\frac{dt_{i}}{\sqrt{2\pi}}  = \frac{1}{\sqrt{1-2c\lambda_i}}
\end{split}
\end{equation}
apart from some normalization constants. Thus,
\begin{equation}
\mathbb{P}\left(\sum_{i>g_1}\lambda_i t_i^2 > u - \frac{\lambda_1 a}{\tilde{\mu}_{min}}\right) \leq e^{-c(u-\lambda_1a/\tilde{\mu}_{min})} \prod_{i>g_1} \frac{1}{\sqrt{1-2c\lambda_i}}
\end{equation}
Also note that the above product is necessarily well behaved if $c<1/2\lambda_{g_1+1}$, hence we impose this constraint on $c$. Hence now we can obtain an upper bound on $\mathbb{P}_u(||\bar{\varphi}||_2^2 < a)\equiv \frac{\mathbb{P}(||\bar{\varphi}||_2^2 < a, \mathcal{Q}>u)}{\mathbb{P}(\mathcal{Q}>u)}$ since we already have a lower bound on $\mathbb{P}(\mathcal{Q}>u)$. We take $c = (\lambda_1^{-1} + \lambda_{g_1+1}^{-1})/4$ and substitute in the above.
\begin{equation}
\exp\left[-c(u-\lambda_1a/\tilde{\mu}_{min})\right] = \exp\left[-\left(\frac{1}{\lambda_1} + \frac{1}{\lambda_{g_1+1}}\right)\frac{u}{4}\right]\exp\left[\frac{a(\lambda_1 + \lambda_{g_1+1})}{4\lambda_{g_1+1}\tilde{\mu}_{min}}\right]
\end{equation}
Thus,
\begin{equation}
\mathbb{P}_{u}((||\bar{\varphi}||_2^2 < a) \leq C_2(\alpha)\exp\left[-\left(\frac{1}{\lambda_1} + \frac{1}{\lambda_{g_1+1}}\right)\frac{u}{4}\right],
\end{equation}
where,
\begin{equation}
C_2(\alpha) = \exp\left[\frac{a(\lambda_1 + \lambda_{g_1+1})}{4\lambda_{g_1+1}\tilde{\mu}_{min}}\right]\prod_{i>g_1}\frac{1}{1-(\lambda_1^{-1} + \lambda_{g_1+1}^{-1})\lambda_i/2}
\end{equation}
Using the bound on $\mathbb{P}(\mathcal{Q}>u)$, we finally obtain:
\begin{equation}
\mathbb{P}_u(||\bar{\varphi}||_2^2 < a)\leq \frac{C_2(\alpha)}{C_1(\alpha)}\frac{1}{u^{g_1/2-1}}\exp\left[-\left(\frac{1}{\lambda_{g_1+1}} - \frac{1}{\lambda_1}\right)\frac{u}{4}\right]
\end{equation}
Next we estimate the conditional probability $\mathbb{P}_u(||\delta\varphi||_2^2>\epsilon a)$. Again noting the fact that $\mathbb{P}(\mathcal{Q}>u)\leq \mathbb{P}\left(\sum_{i\geq 1}{\lambda_i t_i^2}>u\right)$, we write the joint probability measure of $t_{i\notin [1,g_1]}$ as:
\begin{equation}
d\mathbb{P}\left(t_{i\notin [1,g_1]},\sum_{i}{\lambda_it_i^2 >u}\right)\leq d\mathbb{P}\left(t_{i\notin [1,g_1]},\sum_{i\geq 1}{\lambda_it_i^2 >u}\right)
\end{equation}
Also we note that,
\begin{equation}
\mathbb{P}\left(\sum_{i\geq 1}{\lambda_it_i^2 >u}\right) = \mathbb{P} \left(\lambda_1\sum_{i=1}^{g_1}{t_i^2}>u-\sum_{i>g_1}{\lambda_it_i^2}\right),
\end{equation}
and hence we can write the following relation:
\begin{equation}
d\mathbb{P}\left(t_{i\notin [1,g_1]},\sum_{i}{\lambda_it_i^2 >u}\right)\leq \mathbb{P} \left(\sum_{i=1}^{g_1}{t_i^2}>\frac{u}{\lambda_1}-\sum_{i>g_1}{\frac{\lambda_i}{\lambda_1}t_i^2}\right)\prod_{i\notin[1,g_1]}e^{-t_i^2/2}\frac{dt_i}{\sqrt{2\pi}}
\end{equation}
Evaluation of $d\mathbb{P}\left(\sum_{i=1}^{g_1}{t_i^2}\right)$ follows the same procedure as we followed earlier - writing the characteristic function and taking the inverse Fourier transform. We obtain the following pdf:
\begin{equation}
\rho(v) =\frac{2^{(g_1-1)/2}(-1)^{g_1/2}}{2(1\cdot 3\cdot 5\cdot (g_1-2)\sqrt{\pi})} \int {H(v) v^{\frac{g_1}{2}-1}}e^{-v}dv
\end{equation}
where $H(v)$ is the Heavyside step function. Denoting the constant term on the left of the integral as $S$ we obtain:
\begin{equation}
\mathbb{P} \left(\sum_{i=1}^{g_1}{t_i^2}>\frac{u}{\lambda_1}-\sum_{i>g_1}{\frac{\lambda_i}{\lambda_1}t_i^2}\right) = S\int_{\frac{u}{\lambda_1}-\sum_{i>g_1}{\frac{\lambda_i}{\lambda_1}t_i^2}}^{\infty}{H(v) v^{\frac{g_1}{2}-1}e^{-v}dv}
\end{equation}
Performing a change of variables $v\rightarrow v-\sum_{i>g_1}{\lambda_it_i^2/\lambda_1}$ and using the fact that $H(v)v^{\frac{g_1}{2}-1}$ is an increasing function of $v$ for $g>1$ we obtain:
\begin{equation}
\mathbb{P} \left(\sum_{i=1}^{g_1}{t_i^2}>\frac{u}{\lambda_1}-\sum_{i>g_1}{\frac{\lambda_i}{\lambda_1}t_i^2}\right)\leq S\int_{u/\lambda_1}^{\infty}{v^{\frac{g_1}{2}-1}e^{-v}dv \prod_{i>g_1}e^{\lambda_it_i^2/\lambda_1}}
\end{equation}
We perform integration by parts and note that the subsequent terms are of lesser order than the first one (can be neglected taking the limit $u\rightarrow +\infty$) and obtain:
\begin{equation}
\int_{u/\lambda_1}^{\infty}{v^{\frac{g_1}{2}-1}e^{-v}dv}\sim \left(\frac{u}{\lambda_1}\right)^{\frac{g_1}{2}-1}e^{-u/\lambda_1}
\end{equation}
Thus for every $\alpha>0$, there is a $v_1>0$ such that for every $u>v_1$,
\begin{equation}
\mathbb{P} \left(\sum_{i=1}^{g_1}{t_i^2}>\frac{u}{\lambda_1}-\sum_{i>g_1}{\frac{\lambda_i}{\lambda_1}t_i^2}\right)\leq S(1+\alpha){\left(\frac{u}{\lambda_1}\right)^{\frac{g_1}{2}-1}e^{-u/\lambda_1} \prod_{i>g_1}e^{\lambda_it_i^2/2\lambda_1}}
\end{equation}
For $0<\alpha<1$, it follows that for $u$ large enough,
\begin{equation}
d\mathbb{P}_u(\{t_i\notin[1,g_1]\})\leq C_3(\alpha)\left(\prod_{i<0}\frac{e^{-t_i^2/2}}{\sqrt{2\pi}}dt_i\right)\left(\prod_{i>g_1}\frac{e^{-(1-\lambda_i/\lambda_1)t_i^2/2}}{\sqrt{2\pi}}dt_i\right),
\end{equation}
where
\begin{equation}
C_3(\alpha) = \frac{(1+\alpha)S}{C_1(\alpha)\lambda_1^{g_1/2-1}}
\end{equation}
Next we estimate $\mathbb{E}_{u}[e^{c||\delta\varphi||_2^2}]$ for some $c>0$.
\begin{equation}
\mathbb{E}_{u}[e^{c||\delta\varphi||_2^2}] \leq C_3(\alpha) \int{e^{c\sum_{i,j\notin[1,g_1]}{t_i\langle\lambda_i|C|\lambda_j\rangle}t_j}}\left(\prod_{i<0}{e^{-t_i^2/2}\frac{dt_i}{\sqrt{2\pi}}}\right)\left(\prod_{i>g_1}{e^{-(1-\lambda_i/\lambda_1)t_i^2/2}\frac{dt_i}{\sqrt{2\pi}}}\right)
\end{equation}
We can write the above in a compact form as follows:
\begin{equation}
\mathbb{E}_{u}[e^{c||\delta\varphi||_2^2}] \leq C_3(\alpha) \int{e^{-\sum_{i,j\notin[1,g_1]}{t_iG_{ij}t_j}}\prod_{i,j\notin[1,g_1]}\frac{dt_i}{\sqrt{2\pi}}}
\end{equation}
where $
hat{G}$ is the matrix below
\begin{equation}
G_{ij} = diag(\min\{1,1-\lambda_i/\lambda_1\}) - 2c\langle\lambda_i|\hat{C}|\lambda_j\rangle
\end{equation}
For the above Gaussian integral to exist, the matrix $\hat{G}$ must be strictly positive-definite. When we perform the Gaussian integration we end up with an infinite determinant i.e. an infinite product of the eigenvalues which needs to be finite. Thus the matrix diag $(max\{0,\lambda_i/\lambda_1\}) + 2c\langle\lambda_i|\hat{C}|\lambda_j\rangle$ needs to be a trace class. From which it follows that $\hat{M}$ and $\hat{C}$ must be trace class operators. Noticing that $||\hat{C}|| = \mu_1$ and $\min\{1,1-\lambda_i/\lambda_1\}\geq1-\lambda_{g_1+1}/\lambda_1$ for $i\notin[1,g_1]$, then this condition is satisfied for $c<(1-\lambda_{g1+1}/\lambda_1)/2\mu_1$. Thus there exists a $c>0$ such that the Gaussian integral exists-which we call as $C_4(c)$. Thus,
\begin{equation}
\mathbb{E}_{u}[e^{c||\delta\varphi||_2^2}] \leq C_3(\alpha) C_4(c)
\end{equation}
Using exponential Markov inequality, we obtain:
\begin{equation}
\mathbb{P}_u(||\delta\varphi||_2^2>\epsilon a)\leq e^{-\epsilon ca}\mathbb{E}_{u}[e^{c||\delta\varphi||_2^2}]\leq e^{-\epsilon ca} C_3(\alpha) C_4(c)
\end{equation}
Thus plugging both estimates into our original equation and taking the limit $u\rightarrow\infty$ we obtain:
\begin{equation}
\lim_{u\rightarrow\infty}\mathbb{P}_u(||\delta\varphi||_2^2>\epsilon||\varphi||_2^2)\leq e^{-\epsilon ca} C_3(\alpha) C_4(c)
\end{equation}
In the above $a$ can be arbitrarily large and leads to the conclusion that $\lim_{u\rightarrow\infty}P_u(||\delta\varphi||_2^2>\epsilon||\varphi||_2^2) = 0$.\\

Next we consider the case when $g_1=1$ case. We can not use the same approach since $v^{-1/2}$ is not an increasing function of $v$. Let us write $\sigma = \sum_{i>1}{\lambda_it_i^2}$. For a fixed $0<\epsilon<1$ we consider two domains $\sigma < (1-\epsilon)u$ and $\sigma\geq (1-\epsilon)u$. \\

\textbf{Domain 1}\\

We start with the following p.d.f:
\begin{equation}
\rho(v) = S \int{H(v) v^{-\frac{1}{2}}e^{-v/2}dv}
\end{equation}
where $H(v)$ is the Heavyside step function and $S$ is the constant defined earlier. Hence,
\begin{equation}
\mathbb{P} \left({t_1^2}>\frac{u-\sigma}{\lambda_1}\right) = S\int_{\frac{u-\sigma}{\lambda_1}}^{\infty}{H(v) v^{-\frac{1}{2}}e^{-v/2}dv}
\end{equation}
Performing a change of variables $v\rightarrow v-\sigma/\lambda_1$ we obtain:
\begin{equation}
\mathbb{P} \left(t_1^2>\frac{u-\sigma}{\lambda_1}\right) = \int_{u/\lambda_1}^{\infty}{H(v-\sigma/\lambda_1) (v-\sigma/\lambda_1)^{-\frac{1}{2}}e^{-v/2}e^{\sigma/2\lambda_1}dv}
\end{equation}
We note that $(v-\sigma/\lambda_1)^{-\frac{1}{2}}$ is increasing in $\sigma$ and decreasing in $v$ and thus can be bounded by $\left(\frac{\epsilon u}{\lambda_1}\right)^{-1/2}$. Using asymptotics for the integral as we have done often earlier one obtains the following relation:
\begin{equation}
\mathbb{P} \left(t_1^2>\frac{u-\sigma}{\lambda_1}\right) \leq \frac{2\sqrt{\lambda_1}}{\sqrt{\epsilon u}}e^{-u/2\lambda_1}\prod_{i>g_1}e^{\lambda_it_i^2/2\lambda_1}
\end{equation}
Thus for $u$ large enough one has a similar relation as earlier:
\begin{equation}
d\mathbb{P}_u(\{t_i\notin[1,g_1]\})\leq C_3(\alpha)\left(\prod_{i<0}\frac{e^{-t_i^2/2}}{\sqrt{2\pi}}dt_i\right)\left(\prod_{i>g_1}\frac{e^{-(1-\lambda_i/\lambda_1)t_i^2/2}}{\sqrt{2\pi}}dt_i\right),
\end{equation}
where
\begin{equation}
C_3(\alpha) = \frac{2}{C_1(\alpha)}\sqrt{\frac{\lambda_1}{\epsilon}}
\end{equation}
and similarly
\begin{equation}
E_{u}[e^{c||\delta\varphi||_2^2}] \leq C_3(\alpha) C_4(c),
\end{equation}
where $C_4(c)$ is the same as defined earlier. The rest of the proof is the same.\\

\textbf{Domain 2}\\
We can use the following trivial bound:
\begin{equation}
d\mathbb{P}\left(\{t_{i\neq 1}\},\sum_i{\lambda_it_i^2 >u}\right)\leq \prod_{i\neq 1}\frac{e^{-t_i^2/2}}{\sqrt{2\pi}}dt_i
\end{equation}
We make use of the following lower bound 
\begin{equation}
\mathbb{P}(\mathcal{Q}>u)\geq C_1(\alpha)u^{g_1/2-1}e^{-u/2\lambda_1}
\end{equation}
to obtain the following relation for the conditional expectation:
\begin{equation}
\mathbb{E}_u[e^{c||\delta\varphi||_2^2}\textbf{1}_{\sigma\geq(1-\epsilon)u}]\leq \frac{\sqrt{u}e^{u2/\lambda_1}}{C_1(\alpha)}\mathbb{E}[e^{c||\delta\varphi||_2^2}\textbf{1}_{\sigma\geq(1-\epsilon)u}]
\end{equation}
In the above $\textbf{1}_{\sigma\geq(1-\epsilon)u}$ is the indicator function to make use of the fact that we are in the second domain. We use H\"{o}lders inequality to obtain the following relation:
\begin{equation}
\mathbb{E}[e^{c||\delta\varphi||_2^2}\textbf{1}_{\sigma\geq(1-\epsilon)u}]\leq\mathbb{E}[e^{c||\delta\varphi||_2^2}]^{1/q}\mathbb{E}[\textbf{1}_{\sigma\geq(1-\epsilon)u}]^{1/p} = \mathbb{E}[e^{c||\delta\varphi||_2^2}]^{1/q} \mathbb{P}[\sigma\geq(1-\epsilon)u]^{1/p}
\end{equation}
Hence using the bound on $\mathbb{E}[e^{c||\delta\varphi||_2^2}]$ we obtained earlier to write the following:
\begin{equation}
\mathbb{E}_u[e^{c||\delta\varphi||_2^2}\textbf{1}_{\sigma\geq(1-\epsilon)u}]\leq \frac{C_4(qc)^{1/q}}{C_1(\alpha)}\sqrt{u}e^{u/2\lambda_1}\mathbb{P}[\sigma\geq(1-\epsilon)u]^{1/p}
\end{equation}
In the above $p$ and $q$ are H\'{o}lder equivalents i.e $1/p + 1/q = 1$. Using asymptotics again we can evaluate $\mathbb{P}[\sigma\geq(1-\epsilon)u]$ (exactly like how we calculated the pdf for $\mathcal{Q}$ earlier:
\begin{equation}
\mathbb{P}[\sigma\geq(1-\epsilon)u] \leq (1+\alpha)C_5(\alpha)u^{g_2/2-1}\exp\left(-\frac{(1-\epsilon)u}{2\lambda_2}\right)
\end{equation}
Hence for $u$ large enough we obtain:
\begin{equation}
\mathbb{E}_u[e^{c||\delta\varphi||_2^2}\textbf{1}_{\sigma\geq(1-\epsilon)u}]\leq (1+\alpha)\frac{C_5(\alpha)C_4(qc)^{1/q}}{C_1(\alpha)} u^{1/2+g_2/2p - 1/p}\exp\left(-\gamma(\epsilon,p)u\right)
\end{equation}
where
\begin{equation}
\gamma(\epsilon,p)= \frac{1}{2}\left(\frac{1-\epsilon}{p\lambda_2}-\frac{1}{\lambda_1}\right)
\end{equation}
We need $\gamma(\epsilon,p)>0$ and it can be fixed by the following choice: $0<\epsilon<1-\lambda_2/\lambda_1$, $1<p<(1-\epsilon)\lambda_1/\lambda_2$. For the Gaussian integral to exist we should have $c<(1-\lambda_{g_1+1}/\lambda_1)$. Thus for $u$ large enough one has:
\begin{equation}
\mathbb{E}_u[e^{c||\delta\varphi||_2^2}\textbf{1}_{\sigma\geq(1-\epsilon)u}]\leq C_6(\alpha)u^{1/2+g_2/2p-1/p}e^{-\gamma(\epsilon,p)u}
\end{equation}
We recall that $C_4(c)$ was the Gaussian integral we came across earlier. Since $q>1$ one has $C_4(c)\leq C_4(qc)$. Hence we finally use exponential Markov inequality and combine our earlier result of the first domain to obtain the following:
\begin{equation}
\mathbb{P}_u (||\delta\varphi||_2^2 > \epsilon a)\leq e^{-\epsilon ca} [C_3(\alpha)C_4(c) + C_6(\alpha)u^{1/2 +g_2/2p-1/p}e^{-\gamma(\epsilon,p)u}]
\end{equation}
\subsection{Proof when $\lambda_{g_1+1}=0$}
There is no modification until we obtain the following lower bound on $\mathbb{P}(\mathcal{Q}>u)$:
\begin{equation}
\mathbb{P}(\mathcal{Q}>u)\geq C_1(\alpha) u^{g_1/2-1}\exp\left(\frac{-u}{2\lambda_1}\right)
\end{equation}
Then we estimate the conditional probability $\mathbb{P}_u(||\bar{\varphi}||_2^2<a)$ and obtain the following relation:
\begin{equation}
\mathbb{P}(||\bar{\varphi}||_2^2<a,\mathcal{Q}>u)\leq \mathbb{P}\left(\bar{\mu}_{min}\sum_{i=1}^{g_1}{t_i^2}<a, \sum_{i\geq 1}{\lambda_i t_i^2}>u\right)
\end{equation}
Since $\lambda_{g_1+1}$ is zero and infinitely degenerate, we have:
\begin{equation}
\mathbb{P}\left(\bar{\mu}_{min}\sum_{i=1}^{g_1}{t_i^2}<a, \sum_{i\geq 1}{\lambda_i t_i^2}>u\right) = \mathbb{P}\left(\bar{\mu}_{min}\sum_{i=1}^{g_1}{ t_i^2}<a, \sum_{i=1}^{g_1}{\lambda_i t_i^2}>u\right)
\end{equation}
Thus we need to evaluate the following:
\begin{equation}
\mathbb{P}\left(\frac{u}{\lambda_1} < \sum_{i=1}^{g_1}{t_i^2} < \frac{a}{\bar{\mu}_{min}}\right) 
\end{equation}
We note that the above is zero for $u>\lambda_1 a/{\bar{\mu}_{min}}$. So $\mathbb{P}(||\bar{\varphi}||_2^2<a, \mathcal{Q}>u) = 0$ for $u>\lambda_1 a/{\bar{\mu}_{min}}$. Hence we have the result that for every $a>0$, for $u>\lambda_1 a/{\bar{\mu}_{min}}$, $\mathbb{P}_u(||\bar{\varphi}||_2^2<a) = 0$.\\

We next move on to the part where we estimate $\mathbb{P}_{u}(||\delta\varphi||_2^2 > \epsilon a)$. We write the following conditional probability measure:
\begin{equation}
d\mathbb{P}\left(t_{i\neq [1,g_1]},\mathcal{Q}>u\right)\leq d\mathbb{P}\left(t_{i\neq [1,g_1]},\sum_{i\geq 1}{\lambda_i t_i^2}>u\right)
\end{equation}
Since $\lambda_{g_1+1}$ is zero and infinitely degenerate, we have:
\begin{equation}
d\mathbb{P}\left(t_{i\neq [1,g_1]},\sum_{i\geq 1}{\lambda_i t_i^2}>u\right) = d\mathbb{P}\left(t_{i\neq [1,g_1]},\sum_{i=1}^{g_1}{\lambda_i t_i^2}>u\right) = \mathbb{P}\left(\sum_{i=1}^{g_1}{t_i^2}>\frac{u}{\lambda_1}\right)\prod_{i\neq [1,g_1]}{\frac{e^{-t_i^2/2}}{\sqrt{2\pi}}dt_i}
\end{equation}
For large $u$ one has (also since $H(v)=1$ in the considered domain (see below)):
\begin{equation}
\mathbb{P}\left(\sum_{i=1}^{g_1}{t_i^2}>\frac{u}{\lambda_1}\right) = S\int_{u/\lambda_1}^{\infty}{{H(v) v^{g_1/2-1}}e^{-v/2}dv}\sim S\left(\frac{u}{\lambda_1}\right)^{g_1/2-1}e^{-u/2\lambda_1}
\end{equation}
where $S$ is the same constant as defined in the earlier case. Hence for every $\alpha>0$, there is a $v_1>0$ such that for every $v>v_1$,
\begin{equation}
\mathbb{P}(||\bar{\varphi}||_2^2<a,\mathcal{Q}>u)\leq S(1+\alpha)\left(\frac{u}{\lambda_1}\right)^{g_1/2-1}e^{-u/2\lambda_1}\prod_{i\neq [1,g_1]}{\frac{e^{-t_i^2/2}}{2\sqrt{2\pi}}dt_i}
\end{equation}
The conditional probability (for $u>\max\{v_0,v_1\}$) will be:
\begin{equation}
d\mathbb{P}_u (\{t_{i\neq [1,g_1]}\})\leq C_3(\alpha)\prod_{i\neq [1,g_1]}{\frac{e^{-t_i^2/2}}{\sqrt{2\pi}}dt_i},
\end{equation}
where $C_3(\alpha)$ is the same as defined in the case of $\lambda_{g_1+1}\geq 0$. Thus one obtains for large $u$, the following: 
\begin{equation}
\mathbb{E}_u[\exp(c||\delta\varphi||_2^2)]\leq C_3(\alpha)C_4(c),
\end{equation}
where $C_4(c)$ is a Gaussian integral whose existence is ensured if we choose $c<1/2\mu_1$ where $||\hat{C}||=\mu_1$ and $\hat{C}$ being trace class.

\clearpage
\section{Proof of proposition 3}
Let $\mathcal{H}$ be  a separable Hilbert space and $\{|\mu_i\rangle\}$ be an orthonormal basis of $\mathcal{H}$. $\hat{C}$ is a positive trace-class operator acting on $\mathcal{H}$ with the following eigenvalue expansion:
\begin{equation}
\hat{C}=\sum_i \mu_i |\mu_i\rangle\langle \mu_i|
\end{equation}
In the above all the eigenvalues $\mu_i$ are strictly positive and hence kernel of $\hat{C}$ or $\hat{C}^{1/2}$ is zero. Let $|\lambda\rangle\in\mathcal{H}$ be an eigenvector of $M$ with eigenvalue $\lambda$. One can write the following relation:
\begin{equation}
\hat{M}|\lambda\rangle = \hat{C}^{1/2}\hat{O}\hat{C}^{1/2}|\lambda\rangle = \lambda|\lambda\rangle
\end{equation}
Applying $\hat{C}^{1/2}$ to both sides of the above equation we obtain:
\begin{equation}
\hat{C}\hat{O}\hat{C}^{1/2}|\lambda\rangle = \lambda \hat{C}^{1/2}|\lambda\rangle
\end{equation}
The norm of the state $\hat{C}^{1/2}|\lambda\rangle$ is $\langle\lambda|\hat{C}|\lambda\rangle \leq \mu_1$. Hence the state $\hat{C}^{1/2}|\lambda\rangle$ exists and this state belongs to $\mathcal{D}(\hat{C}^{-1/2})$. Thus if $\lambda$ is also an eigenvalue of $\hat{C}\hat{O}$ with eigenvector $\hat{C}^{1/2}|\lambda\rangle$. Thus the spectrum of $\hat{M}$ is a subset of spectrum of restriction of $\hat{C}\hat{O}$ to $\mathcal{D}(\hat{C}^{-1/2})$.\\

Now let $|\phi\rangle \in\mathcal{D}(\hat{C}^{-1/2})$ be an eigenstate of $\hat{C}\hat{O}$ with eigenvalue $\lambda$. We have the following eigenvalue equation:
\begin{equation}
\hat{C}\hat{O}|\phi\rangle = \lambda|\phi\rangle
\end{equation}
The state $|\lambda\rangle = \hat{C}^{-1/2}|\phi\rangle$ exists (that was the assumption earlier) and hence we can write $|\phi\rangle = \hat{C}^{1/2}|\lambda\rangle$. Inserting this in the previous eigenvalue equation, one obtains:
\begin{equation}
\hat{C}^{1/2}(M|\lambda\rangle - \lambda |\lambda\rangle) = 0
\end{equation}
Since kernel $\hat{C}^{1/2} = \{0\}$ we have $M|\lambda\rangle = \lambda|\lambda\rangle$. Thus $\lambda$ is also an eigenvalue of $\hat{M}$ and the spectrum of restricition of $\hat{C}\hat{O}$ to $\mathcal{D}(\hat{C}^{-1/2})$ is subset of spectrum of $M$. Thus the proof of proposition is complete.

\clearpage
\section{Calculations for local helicity}
Let us consider $\bar{c}_1(\bar{x})\cdot$ curl $v(0)$. It can be written as follows for the above form of correlation function: 
\begin{equation}
\begin{split}
\left(\frac{E}{3}xf'(x) + \frac{2E}{3}f(x) - \frac{E}{3}xf'(x)\frac{x_1x_1}{x^2}\right) |\nabla\times \bar{v}(0)|_1 \\&- \left(\frac{E}{3}xf'(x)\frac{x_1x_2}{x^2}\right)|\nabla\times \bar{v}(0)|_2\\& - \left(\frac{E}{3}xf'(x)\frac{x_1x_3}{x^2}\right)|\nabla\times \bar{v}(0)|_3
\end{split}
\end{equation}
Similarly we consider the other components $c_2(x)\cdot$ curl $\bar{v}(0)$ and $c_3(x)\cdot$ curl $\bar{v}(0)$ also. Combining, we can write the resultant vector in the following form:
\begin{equation}
\frac{E}{3}(2f(x) + xf'(x))(\nabla\times \bar{v}(0)) - \frac{E}{3}\bar{x}f'(x)\left(\frac{\bar{x}}{x}\cdot(\nabla\times \bar{v}(0))\right)
\end{equation}
Now let us consider $\bar{v}(0)\cdot$ curl $\bar{c}_1(\bar{x})$. For the given form of correlation function, it can be expanded as:
\begin{equation}
v_1(0)\left(\partial_2 C_{13} - \partial_3 C_{12}\right) + v_2(0)\left(\partial_3 C_{11} - \partial_1 C_{13}\right) + v_3(0)\left(\partial_1 C_{12} - \partial_2 C_{11}\right)
\end{equation}
\begin{equation}
\begin{split}
\left(\partial_2 C_{13} - \partial_3 C_{12}\right) &= -\frac{E}{3}x_1x_3\partial_2 (xf'(x)/x^2) + \frac{E}{3}x_1x_2\partial_3 (xf'(x)/x^2)\\
&= \frac{E}{3}x_1x_3f'(x)\frac{x_2}{x^3} - \frac{E}{3x}x_1x_3x_2\frac{f''(x)}{x} -\frac{E}{3}x_1x_2f'(x)\frac{x_3}{x^3} + \frac{E}{3x}x_1x_3x_2\frac{f''(x)}{x}\\
&= 0 
\end{split}
\end{equation}

\begin{equation}
\begin{split}
\left(\partial_3 C_{11} - \partial_1 C_{13}\right) &= \frac{2E}{3}\partial_3 f(x) + \frac{E}{3}\partial_3 (xf'(x)) - \frac{E}{3} x_1x_1\partial_3 (f'(x)/x) + \frac{E}{3}x_3\partial_1 (f'(x)x_1/x)\\
&= \frac{2E}{3x}x_3f'(x) + \frac{E}{3}x_3f''(x) + \frac{E}{3x}x_3f'(x) - \frac{E}{3x^2}x_1x_1x_3f''(x) - \frac{E}{3x^3}x_1x_1f'(x)x_3 \\
&+ \frac{E}{3x}x_3f'(x) + \frac{E}{3x^2}x_3x_1x_1f''(x) + \frac{E}{3x^3}x_3x_1x_1f'(x)\\
&= \frac{4E}{3x}x_3f'(x) + \frac{E}{3}x_3f''(x)
\end{split}
\end{equation}
Similarly we obtain:
\begin{equation}
\begin{split}
\left(\partial_1 C_{12} - \partial_2 C_{11}\right) = -\frac{4E}{3x}x_2f'(x) - \frac{E}{3}x_2f''(x)
\end{split}
\end{equation}
and thus:
\begin{equation}
\begin{split}
\bar{v}(0)\cdot (\nabla\times \bar{c}_1(\bar{x})) = v_2(0) \left(\frac{4E}{3x}x_3f'(x) + \frac{E}{3}x_3f''(x)\right) - v_3(0) \left(\frac{4E}{3x}x_2f'(x) + \frac{E}{3}x_2f''(x)\right)
\end{split}
\end{equation}
Taking into account the other two components also, one can write the resultant vector in the following compact manner:
\begin{equation}
-\frac{E}{3}\left(4f'(x) + xf''(x)\right)\left(\frac{\bar{x}}{x}\times \bar{v}(0)\right)
\end{equation}
The eigenvalue equation can be written as follows after combining all three vector components:
\begin{equation}
\begin{split}
2v\bar{v}(\bar{x}) = \frac{E}{3}(2f(x) + xf'(x))(\nabla\times \bar{v}(0)) - \frac{E}{3}\bar{x}f'(x)\left(\frac{\bar{x}}{x}\cdot\nabla\times \bar{v}(0)\right) + \frac{E}{3}\left(4f'(x) + xf''(x)\right)\left(\frac{\bar{x}}{x}\times \bar{v}(0)\right)
\end{split}
\end{equation}
We next take the curl of the above equation. Let us first take the curl of $\frac{E}{3}(2f(x) + xf'(x))(\nabla\times \bar{v}(0))$. We obtain the following for the first component of the curl:
\begin{equation}
\begin{split}
|\nabla\times \bar{v}(0)|_3\partial_2 (2f(x) + xf'(x)) - |\nabla\times \bar{v}(0)|_2\partial_3 (2f(x) + xf'(x))\\ = |\nabla\times \bar{v}(0)|_3 \left(\frac{3x_2}{x}f'(x) + x_2f''(x)\right) - |\nabla\times \bar{v}(0)|_2 \left(\frac{3x_3}{x}f'(x) + x_3f''(x)\right)
\end{split}
\end{equation}
Similarly combining the other components, the curl of $\frac{E}{3}(2f(x) + xf'(x))(\nabla\times \bar{v}(0))$ can be written as follows:
\begin{equation}
\frac{E}{3}[3f'(x) + xf''(x)] \left(\frac{\bar{x}}{x}\times (\nabla\times \bar{v}(0))\right)
\end{equation}
Next we consider the curl of $- \frac{E}{3}\bar{x}f'(x)\left(\frac{\bar{x}}{x}\cdot\nabla\times \bar{v}(0)\right)$ and obtain the following:
\begin{equation}
-\frac{E}{3x}x_3f'(x) |\nabla\times \bar{v}(0)|_2 + \frac{E}{3x}x_2f'(x) |\nabla\times \bar{v}(0)|_3
\end{equation}
Combining, the curl of $\left[\frac{E}{3}(2f(x) + xf'(x))(\nabla\times \bar{v}(0)) - \frac{E}{3}\bar{x}f'(x)\left(\frac{\bar{x}}{x}\cdot\nabla\times \bar{v}(0)\right)\right]$ can be written as follows:
\begin{equation}
\frac{E}{3}[4f'(x) + xf''(x)] \left(\frac{\bar{x}}{x}\times (\nabla\times \bar{v}(0))\right)
\end{equation}
Now we are just left with the curl of $\frac{E}{3}[4f'(x) + xf''(x)]\left(\frac{\bar{x}}{x}\times \bar{v}(0)\right)$. This can be expanded as follows:
\begin{equation}
\begin{split}
\frac{E}{3}[4f'(x)+xf''(x)][(x_2v_3(0) - x_3v_2(0))\hat{x}_1 + (x_3v_1(0) - x_1v_3(0))\hat{x}_2 + (x_1v_2(0) - x_2v_1(0))\hat{x}_3]
\end{split}
\end{equation}
Writing the first component of the curl of the above:
\begin{equation}
\begin{split}
\partial_2 \left[\frac{E}{3} (4f'(x) + xf''(x)) (x_1v_2(0) - x_2v_1(0))\right] - \partial_3 \left[\frac{E}{3} (4f'(x) + xf''(x)) (x_3v_1(0) - x_1v_3(0))\right]
\end{split}
\end{equation}
The above can be evaluated to obtain:
\begin{equation}
\begin{split}
 x_1v_2(0)\frac{E}{3x} [4f'(x) + x_2 x f'''(x) + f''(x)x_2] \\&- x_2v_1(0)\frac{E}{3x} [4f'(x) + x_2 x f'''(x) + f''(x)x_2] \\&- v_1(0)\frac{E}{3}[4f'(x) + xf''(x)]\\
&+ x_1v_3(0)\frac{E}{3x} [4f'(x) + x_3 x f'''(x) + f''(x)x_3] \\&- x_3v_1(0)\frac{E}{3x} [4f'(x) + x_3 x f'''(x) + f''(x)x_3] \\&- v_1(0)\frac{E}{3}[4f'(x) + xf''(x)]
\end{split}
\end{equation}
Similarly combining other two components also we obtain the following expression:
\begin{equation}
-\frac{2E}{3x}[4f'(x) + xf''(x)] \bar{v}(0) + \frac{E}{3x}[4f'(x) - 4xf''(x) - x^2f'''(x)]\left[\bar{v}(0) - \left(\frac{\bar{x}}{x}\cdot \bar{v}(0)\right)\frac{\bar{x}}{x}\right]
\end{equation}
Combining all of the above, we get the following equation:
\begin{equation}
\begin{split}
2v (\nabla\times \bar{v}(x)) 
 &=\frac{E}{3}[4f'(x) + xf''(x)] \left(\frac{\bar{x}}{x}\times (\nabla\times \bar{v}(0))\right) \\
 &-\frac{2E}{3x}[4f'(x) + xf''(x)] \bar{v}(0) \\
 &+ \frac{E}{3x}[4f'(x) - 4xf''(x) - x^2f'''(x)]\left[\bar{v}(0) 
  \left(\frac{\bar{x}}{x}\cdot \bar{v}(0)\right)\frac{\bar{x}}{x}\right]
\end{split}
\end{equation}

\end{document}